\documentclass[twocolumn]{aastex63}

\usepackage{xspace}
\usepackage{units}
\usepackage{amsmath}
\usepackage{appendix}

\newcommand{\ammo}{\ensuremath{\rm NH_3}\xspace}

\newcommand{\kms}{\ensuremath{\rm km\,s^{-1}}\xspace}

\newcommand{\vlsr}{\ensuremath{V_{\rm lsr}}\xspace}
\newcommand{\cc}{\ensuremath{\rm cm^{-3}}\xspace}

\newcommand{\jybm}{\ensuremath{\rm Jy\,beam^{-1}}\xspace}

\newcommand{\nhtd}{\ensuremath{\rm NH_2D}}
\newcommand{\nhtdoo}{\ensuremath{\rm NH_2D(1_{11}-1_{01})}}

\received{}
\revised{}
\accepted{\today}
\submitjournal{ApJ}

\shorttitle{Almost Complete Freeze-out}
\shortauthors{Caselli et al.}

\begin{document}

\title{The Central 1000 au of a Pre-stellar Core Revealed with ALMA. II. Almost Complete Freeze-out}

\correspondingauthor{Paola Caselli}
\email{caselli@mpe.mpg.de}

\author[0000-0003-1481-7911]{Paola Caselli}
\affiliation{Max-Planck-Institut f\"ur extraterrestrische Physik, Gie{\ss}enbachstrasse 1, 85748 Garching bei M\"unchen, Germany}

\author[0000-0002-3972-1978]{Jaime E. Pineda}
\affiliation{Max-Planck-Institut f\"ur extraterrestrische Physik, Gie{\ss}enbachstrasse 1, 85748 Garching bei M\"unchen, Germany}

\author[0000-0002-9148-1625]{Olli Sipil\"a}
\affiliation{Max-Planck-Institut f\"ur extraterrestrische Physik, Gie{\ss}enbachstrasse 1, 85748 Garching bei M\"unchen, Germany}

\author{Bo Zhao}
\affiliation{Max-Planck-Institut f\"ur extraterrestrische Physik, Gie{\ss}enbachstrasse 1, 85748 Garching bei M\"unchen, Germany}
\affiliation{Department of Physics and Astronomy, McMaster University, 1280 Main St. W, Hamilton, ON L8S 4M1, Canada}

\author{Elena Redaelli}
\affiliation{Max-Planck-Institut f\"ur extraterrestrische Physik, Gie{\ss}enbachstrasse 1, 85748 Garching bei M\"unchen, Germany}

\author{Silvia Spezzano}
\affiliation{Max-Planck-Institut f\"ur extraterrestrische Physik, Gie{\ss}enbachstrasse 1, 85748 Garching bei M\"unchen, Germany}

\author{Maria Jos\'e Maureira}
\affiliation{Max-Planck-Institut f\"ur extraterrestrische Physik, Gie{\ss}enbachstrasse 1, 85748 Garching bei M\"unchen, Germany}

\author{Felipe Alves}
\affiliation{Max-Planck-Institut f\"ur extraterrestrische Physik, Gie{\ss}enbachstrasse 1, 85748 Garching bei M\"unchen, Germany}

\author{Luca Bizzocchi}
\affiliation{Max-Planck-Institut f\"ur extraterrestrische Physik, Gie{\ss}enbachstrasse 1, 85748 Garching bei M\"unchen, Germany}
\affiliation{Dipartimento di Chimica “Giacomo Ciamician”, Universit\`a di Bologna, Via F. Selmi 2, 40126 Bologna, Italy}
\affiliation{Scuola Normale Superiore, Piazza dei Cavalieri 7, 56126 Pisa, Italy}

\author[0000-0001-7491-0048]{Tyler L. Bourke}
\affiliation{SKA Observatory, Jodrell Bank, Lower Withington, Macclesfield SK11 9FT, UK}

\author{Ana Chac\'on-Tanarro}
\affiliation{Observatorio Astron\'omico Nacional (OAG-IGN), Alfonso XII 3, 28014, Madrid, Spain}

\author{Rachel Friesen}
\affiliation{Department of Astronomy \& Astrophysics, University of Toronto, 50 St. George St., Toronto, ON M5S 3H4, Canada}

\author{Daniele Galli}
\affiliation{INAF-Osservatorio Astrofisico di Arcetri, Largo E. Fermi 5, 50125 Firenze, Italy}

\author{Jorma Harju}
\affiliation{Max-Planck-Institut f\"ur extraterrestrische Physik, Gie{\ss}enbachstrasse 1, 85748 Garching bei M\"unchen, Germany}
\affiliation{Department of Physics, P.O. BOX 64, 00014 University of Helsinki, Finland}

\author{Izaskun Jim\'enez-Serra}
\affiliation{Centro de Astrobiolog\'ia (CSIC-INTA), Ctra. de Torrej{\'o}n a Ajalvir km 4, 28850, Torrej{\'o}n de Ardoz, Spain}

\author{Eric Keto}
\affiliation{Harvard-Smithsonian Center for Astrophysics, 160 Garden Street,  Cambridge, MA 02420, USA}

\author{Zhi-Yun Li}
\affiliation{Department of Astronomy, University of Virginia, Charlottesville, VA 22904, USA}

\author{Marco Padovani}
\affiliation{INAF-Osservatorio Astrofisico di Arcetri, Largo E. Fermi 5, 50125 Firenze, Italy}

\author{Anika Schmiedeke}
\affiliation{Max-Planck-Institut f\"ur extraterrestrische Physik, Gie{\ss}enbachstrasse 1, 85748 Garching bei M\"unchen, Germany}

\author{Mario Tafalla}
\affiliation{Observatorio Astron\'omico Nacional (OAG-IGN), Alfonso XII 3, 28014, Madrid, Spain}

\author{Charlotte Vastel}
\affiliation{IRAP, Universit\'e de Toulouse, CNRS, CNES, UPS, Toulouse, France}



\begin{abstract}

Pre-stellar cores represent the initial conditions in the process of star and planet formation. Their low temperatures ($<$10\,K) allow the formation of thick icy dust mantles, which will be partially preserved in the future protoplanetary disks, ultimately affecting the chemical composition of planetary systems. Previous observations have shown that carbon- and oxygen-bearing species, in particular CO, are heavily depleted in pre-stellar cores due to the efficient molecular freeze-out onto the surface of cold dust grains. However, N-bearing species such as NH$_3$ and, in particular, its deuterated isotopologues, appear to maintain high abundances where CO molecules are mainly in solid phase. Thanks to ALMA, we present here the first clear observational evidence of NH$_2$D freeze-out toward the L1544 pre-stellar core, suggestive of the presence of a ``complete-depletion zone'' within a $\simeq$1800\,au radius, in agreement with astrochemical pre-stellar core model predictions. Our state-of-the-art chemical model coupled with a non-LTE radiative transfer code demonstrates  that NH$_2$D becomes mainly incorporated in icy mantles in the central 2000\,au and starts freezing-out already at $\simeq$7000\,au. Radiative transfer effects within the pre-stellar core cause the \nhtdoo \ emission to appear centrally concentrated, with a flattened distribution within the central $\simeq$3000\,au, unlike the 1.3\,mm dust continuum emission which shows a clear peak within the central  $\simeq$1800\,au. This prevented NH$_2$D freeze-out to be detected in previous observations, where the central 1000\,au cannot be spatially resolved.    

\end{abstract}



\section{Introduction} \label{sec:intro}

The initial conditions of the process of star and planet formation are to be found in dense cloud cores \citep[][]{BerginTafalla2007}, especially in the so-called pre-stellar cores, which are gravitationally bound objects \citep[][]{AndreDiFrancesco2014} with central number densities above 10$^5$\,cm$^{-3}$ \citep[][]{KetoCaselli2008} and large deuterium fractions measured especially in N-bearing species \citep[][]{CrapsiCaselli2005}, which preferentially trace the central regions \citep{HilyBlantWalmsley2010,SpezzanoCaselli2017}. Detailed studies of pre-stellar cores in nearby clouds have provided clues on their physical and chemical structure: they are centrally concentrated, with low central temperatures \citep[$\sim$7\,K,][]{Crapsi2007-L1544,PaganiBacmann2007,LaunhardtStutz2013} and evidence of grain growth \citep{Forbrich2015,Chacon2019}.  They display a clear chemical differentiation, where rare CO isotopologue emission maps present a valley within the central $\sim$8500\,au, due to freeze-out onto dust grains \citep[][]{CaselliWalmsley1999}, while N-bearing species such as N$_2$H$^+$, NH$_3$ and their deuterated forms appear to trace well the core center \citep[][]{TafallaMyers2002,Tafalla2004-Cores,Juarez2017,SpezzanoCaselli2017}, defined as the millimeter and sub-millimeter dust continuum emission peak. Although there is evidence of some central depletion of N$_2$H$^+$ \citep[][]{BerginAlves2002}, N-bearing deuterated species have been found to trace well the dust peak \citep[][]{Caselli2002_L1544-ions}. Only recently, high-sensitivity single-dish observations of multiple transitions of N$_2$D$^+$, coupled with astrochemical and non-LTE radiative transfer modeling, have provided evidence of N$_2$D$^+$ central depletion \citep[][]{Redaelli2019}, but the size of the related depletion zone cannot be measured because of the limited angular resolution. The abundance of N$_2$H$^+$ isotopologues is linked to N$_2$, which has a binding energy similar to that of CO \citep[][]{BisschopFraser2006}, although recent studies have shown that the N$_2$ (but not the CO) binding energy is reduced in CO-N$_2$ ice mixtures \citep[][]{NguyenBaouche2018}. The smaller amount of N$_2$ freeze-out compared to CO can be explained if N$_2$ formation proceeds at a slower pace than CO formation, so that most of N$_2$ is still forming from (the more volatile) atomic nitrogen in the dense material when CO has already started to condense onto icy mantles \citep[][]{FlowerPineauDesForets2006,HilyBlantWalmsley2010,PaganiBourgoin2012}. 

No evidence of ammonia freeze-out in starless and pre-stellar cores has been reported yet, despite the fact that the $\ammo$ binding energy is significantly larger than that of CO and N$_2$ and close to that of water molecules \citep[][]{Sipila2019-NH3_no_freezeout}. This is puzzling, as the freeze-out time scale within the central 1000\,au of pre-stellar cores, where the number density is about 10$^6$\,cm$^{-3}$, is only $\sim$1000\,yr, much shorter than dynamical time scales. Furthermore, it appears that the $\ammo$ abundance is slightly {\it increasing} toward the center when viewed with single-dish and interferometers \citep[][]{TafallaMyers2002,Crapsi2007-L1544,Caselli2017-L1544_Infall}. Such an increasing abundance profile for $\ammo$ is at odds with state-of-the-art astrochemical modeling \citep[][]{Sipila2019-NH3_no_freezeout}. \citet[][]{Crapsi2007-L1544} also observed para-$\nhtdoo$ toward the Taurus pre-stellar core L1544 with the IRAM Plateau de Bure Interferometer, finding an integrated intensity map centered on the dust peak, thus ruling out any central depletion of $\nhtd$ when viewed at an angular resolution of 5.8$\arcsec \times 4.5\arcsec$  \citep[986\,au$\times$765\,au, at the distance of 170\,pc \footnote{We note that previous papers on L1544 adopted a distance to the source of 140\,pc or 135\,pc, based on previous distance measurements \citep{Schlafly2014,Galli2018}. Here we adjust previous results to the new distance estimate.};][]{Galli2019}. From this result, it was concluded that no {\it complete depletion zone}\footnote{A {\it complete depletion zone} is defined as a region within a pre-stellar core where, due to freeze-out, species heavier than He have left the gas phase and reside in the icy mantles of dust grains.} \citep[][]{WalmsleyFlower2004} was present in L1544. 

With ALMA 1.3\,mm dust continuum emission observations of L1544, we have recently revealed a compact central region, with mass $\simeq$0.16\,M$_{\odot}$, radius $\simeq$1800\,au and average H$_2$ number density of $\simeq$10$^6$\cc , called the "kernel" \citep{Caselli2019-L1544_ALMA}. This kernel is the stellar system cradle, which we are investigating in detail to shed light on the first steps toward star and protoplanetary disk formation. Here we present the ALMA (12m+ACA) para-\nhtdoo \, mosaic, which reveals for the first time the $\nhtd$ depletion zone toward the L1544 kernel at an angular resolution of 2.5$\arcsec$ (425\,au; Sections 2 and 3). We also demonstrate that the current data are in agreement with our astrochemical model predictions (Section 4), implying that almost all (99.99\%) species heavier than He reside on dust icy mantles within the kernel. Hereafter, p\nhtd \, indicates para-\nhtd . A detailed study of the kernel kinematics will be presented in a forthcoming paper.

\section{Observations}
\subsection{Deuterated Ammonia}
We observed L1544 with ALMA during Cycle 4 (ESO Project ID 2016.1.00240.S, PI Caselli) using the 12 m (Main array) and the 7 m (ACA) arrays.
The 12 m array 3-pointing observations were carried out on 2016 December 22, while the 7 m array single pointing observations were carried out on 2016 October 2. 
The Main array observations included the quasar J0510+1800 for bandpass and phase calibration, while quasar J0423-0120 is observed for flux calibration.
The ACA observations included the quasar J0510+1800 for flux, bandpass, and phase calibration.
The 3-point mosaic observations used a correlator with
the continuum window centered at 100 GHz and a spectral window for pNH$_2$D (1$_{11}$–1$_{01}$) centered at 110.15355 GHz (these data will be presented in a future paper, dedicated to the study of dust properties within the kernel).
The bandwidth and spectral resolution for the pNH$_2$D observations are 58.59 MHz and 30.518 kHz (corresponding to a velocity resolution of 83\,m\,s$^{-1}$), respectively. 
The 12 m and 7 m array data were calibrated using the Common Astronomy Software
Applications package \citep[CASA][]{McMullin2007-CASA} version 4.7.0.
We assigned weights to the datasets with the \verb+statwt+ command before combining them (\verb+concat+ command).
We perform a joint deconvolution of both data sets (ACA and 12 m
array) in CASA (version 5.6.2) using \verb+tclean+ for mosaics, natural weighting, and restored with a circular beam of 2.5\arcsec. 
We use a rest frequency for the imaging of 110.153562\,GHz \citep{Cohen1982,DeLucia1975}.
We use the multi-scale clean technique, with the multi-scale
parameter of [0, 2.5, 7.5, 22.5]\arcsec. 
The final noise level is 7 mJy beam$^{-1}$ per 0.08 km\,s$^{-1}$ channel, which corresponds to a noise level of 0.13 K in the central 45\arcsec \, diameter region of the primary beam corrected data.
The integrated intensity map of p\nhtdoo \, (ACA+12\,m array) is shown in the top panel of Fig.~\ref{fig:L1544_maps} and the full mosaic is shown in Fig.~\ref{fig:L1544_mosaic}. We note that the combined ACA+12\,m data perfectly recover the single dish flux within the IRAM-30m beam (see Figure~\ref{fig:L1544_spectrum}), demonstrating that mosaic pointing observations with ALMA are a powerful technique to carry out detailed studies of pre-stellar cores.

\subsection{1.3\,mm Continuum}
\begin{figure}
    \centering
    \includegraphics[width=0.47\textwidth]{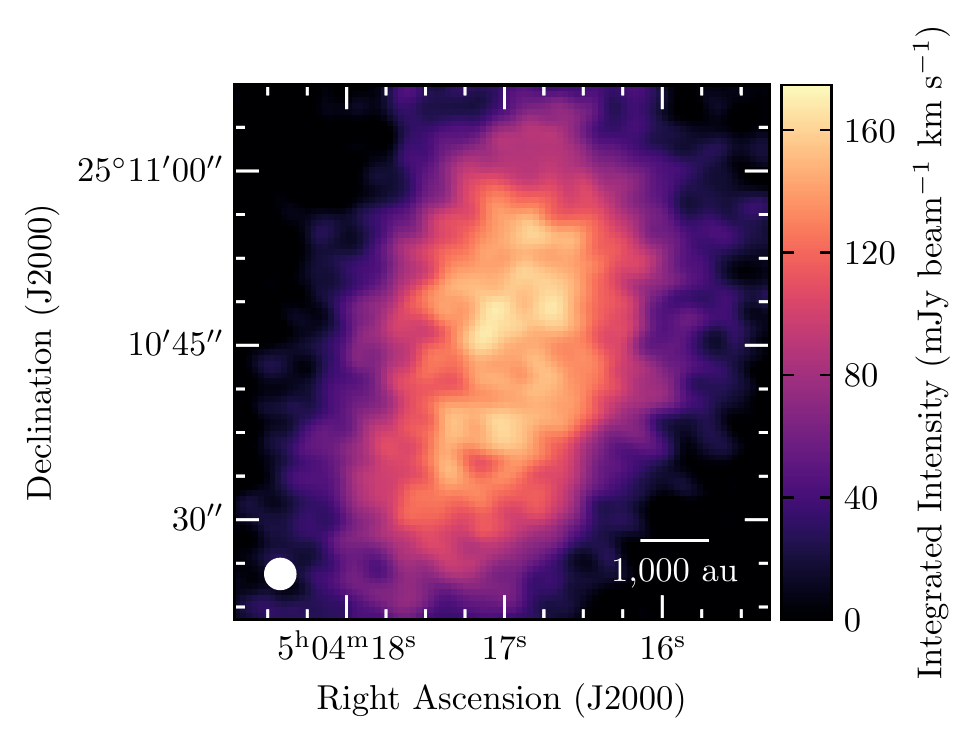}
    \includegraphics[width=0.47\textwidth]{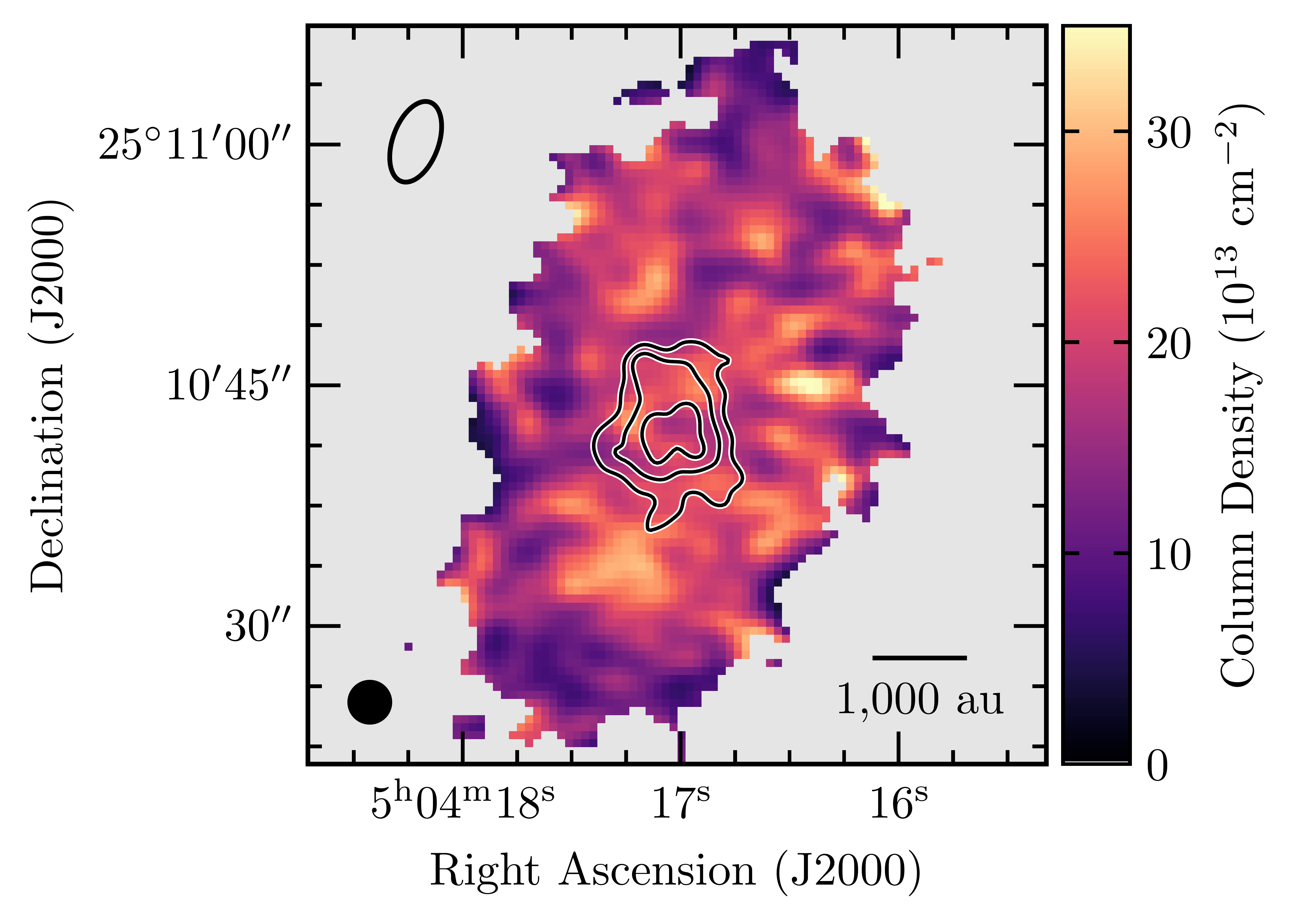}
    \caption{{\sl Top}: Integrated intensity map of the p\nhtdoo \, transition toward L1544. {\sl Bottom}: pNH$_2$D column density map derived from fitting the hyperfine structure. The contours show the 14, 17, and 20 $\times \sigma$ levels ($\sigma$=36\,$\mu$\jybm) of the 1.3\,mm dust continuum emission map presented in \cite{Caselli2019-L1544_ALMA}, which locates the kernel. 
    Scale bar and beam size are shown at the bottom right and left corners, respectively. 
    In the top left of the bottom panel we show the orientation and shape of the ellipse used to define the projected radius in Sect.\,\ref{sec:linefit}  and Fig.\,\ref{fig:column_density_profile}.
    The typical uncertainty on the pNH$_2$D column density is about 10\%.}
    \label{fig:L1544_maps}
\end{figure}

We observed L1544 with ALMA during Cycle 2 (ESO Project ID 2013.1.01195.S, PI Caselli) using both the 12\,m (Main array) and the 7\,m (ACA) arrays. 
The data were already presented in \cite{Caselli2019-L1544_ALMA}, but here we briefly describe the data reduction process.
The 12\,m array observations were carried out on 2014 December 27 and 29, while the 7\,m array observations were carried out on 2014 June 15, July 20 and 29, and August 6 and 11. 
The single pointing observations used a correlator with the continuum centered at 228.973\,GHz and a 2\,GHz bandwidth. 
The 12\,m array data were calibrated using the CASA version 4.2.2, 
while the 7\,m array data were calibrated using CASA version 4.2.1. 
We run the \verb+statwt+ command on both datasets before combination.
We perform a joint deconvolution of both data sets (ACA and 12\,m
array) in CASA (version 5.6.2) using \verb+tclean+ multifrequency synthesis
(mfs) for mosaics, a natural weighting, a taper of 2.1\arcsec$\times$1.45\arcsec \, and PA of 147.2$^{\circ}$, and restored with a circular beam of 2.5\arcsec.
Since the continuum emission is extended, we
use the multi-scale clean technique, with the multi-scale
parameter of [0, 1.25, 3.75, 11.25]\arcsec. The 1.3\,mm dust continuum emission map can be seen in the bottom panel of Fig.\,\ref{fig:L1544_maps}, overlaid onto the pNH$_2$D column density map, which will be presented in Sect.~\ref{sec:linefit}.

\section{Column Density Calculation}
 \label{sec:linefit}
We perform the line fit of p\nhtdoo \, using the \verb+nh2d+ model implemented in \verb+pyspeckit+ \citep{pyspeckit}.
The pNH$_2$D (1$_{11}$-0$_{01}$) line is modeled using the hyperfine structure and relative intensity from the recent laboratory works of \citet{Melosso2020} and \citet{Melosso2021-NH2D_frequencies}.
The model assumes a Gaussian velocity distribution (with centroid velocity and velocity dispersion \vlsr and $\sigma_v$), equal excitation temperatures ($T_{ex}$) for all hyperfine components, and total optical depth ($\tau_0$). 
We perform a fit for the whole cube with all 4 free parameters.
Following the procedure in \cite{Pineda2021-Ion_Neutrals}, we discard all centroid velocity determinations if the uncertainty in the \vlsr is larger than 0.02 \kms, the velocity dispersion determinations if the uncertainty $\sigma_v$ is larger than 0.015\,\kms, and the excitation temperature if the uncertainty is larger than 1\,K.

The column density has been derived following \cite{Mangum-2015}:
\begin{align}
    N({\rm pNH_2D}) = \frac{8\pi \nu^3}{c^3 A_{ul}} \frac{Q(T_{ex})}{g_u e^{-E_u/kT_{ex}}} \nonumber \\
     \times \left[e^{h\nu/kT_{ex}}-1\right]^{-1}\int \tau dv~,
     \label{eq:column_density}
\end{align}
where
$\int \tau dv = \sqrt{2\pi}\ \tau_0 \sigma_v$, 
$\nu$ is the frequency of the transition, 
$g_u$  is the upper level degeneracy, 
$E_u$ is the upper energy level of the transition,
$A_{ul}$ is the Einstein coefficient, and 
$Q(T_{ex})$ is the partition function.
The values for the $E_u$, $g_u$, and $A_{ul}$ are obtained from the Leiden Atomic and Molecular Database \citep[LAMDA;][]{Schoier2005-LAMBDA,vdTak2020}\footnote{\url{https://home.strw.leidenuniv.nl/~moldata/datafiles/p-nh2d.dat}}, 
and the partition function is calculated up to the first 30 energy levels. 
This formula has been used previously  \citep{Daniel2016-NH2D,Harju2017-HMM1_NH2D}. The $T_{ex}$ has been estimated at each position using the hfs fitting within \verb+pyspeckit+, with average uncertainties of 2\% (and a maximum of 5\% at the edge of the map). The error associated with the column density is about 10\%, close to the calibration error.

\begin{figure}
    \centering
    \includegraphics[width=0.4\textwidth]{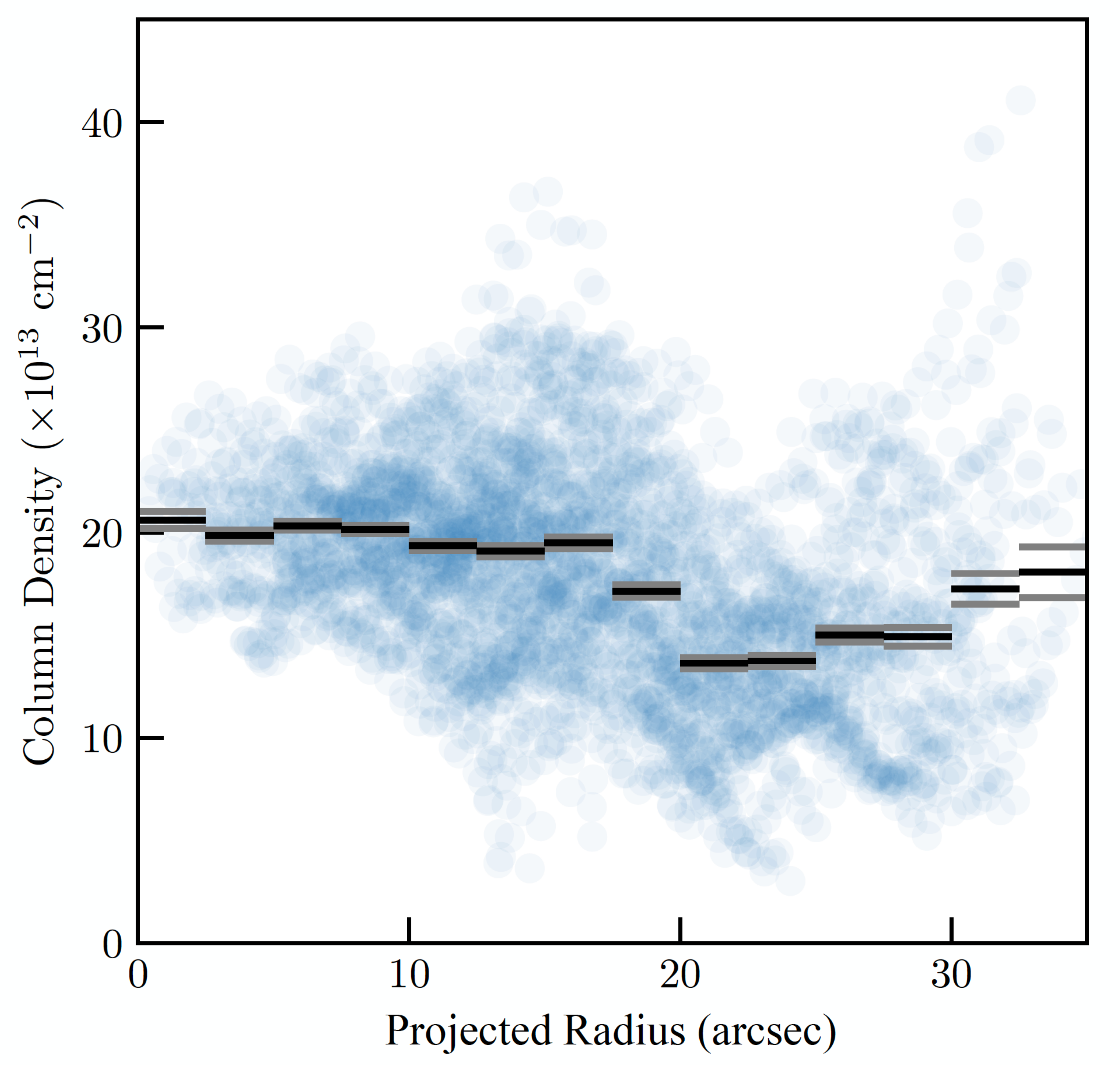}
    \caption{pNH$_2$D column density as a function of projected radius (see main text for its definition). Blue circles correspond to individual measurements across the column density map in the bottom panel of Fig.\ref{fig:L1544_maps}. The black and gray horizontal lines are the mean and corresponding uncertainty within a 2.5\arcsec \, bin. Note the flat column density profile within the central 10\arcsec , where the 1.3\,mm dust continuum map shows a clear peak (see Fig.\,\ref{fig:L1544_maps}, bottom panel).}
    \label{fig:column_density_profile}
\end{figure}

The column density map in Fig.\,\ref{fig:L1544_maps} shows a complex
structure, partly due to noise, without a clear peak toward the 1.3\,mm dust continuum peak, which identifies the kernel \citep{Caselli2019-L1544_ALMA}. This implies that the pNH$_2$D column density does not follow the H$_2$ column density, thus a drop in pNH$_2$D fractional abundance (with respect to H$_2$) within the kernel radius of $\simeq$1800\,au  ($\simeq$10\arcsec ) must be present. The pNH$_2$D column density flattening within the central $\simeq$3000\,au, which includes the L1544 kernel, is well illustrated in Fig.\,\ref{fig:column_density_profile}. The circles represent individual measurements, while the black and gray horizontal lines are the mean and its associated uncertainty in a 2.5\arcsec \, bin.
The projected radius $pr$ is calculated as $pr = \sqrt{r_{maj}^2 + (r_{min} b_{axes}) ^2}~,$
where $r_{maj}$ and $r_{min}$ are the distances along the semi-major and -minor axes, respectively; 
$b_{axes}=22/12$ is the semi-major and -minor axis ratio for the pNH$_2$D emission, where the semi-major axis is along the position angle of 160$^{\circ}$ (measured East from North; see the bottom panel of Fig.\,\ref{fig:L1544_maps} for an illustration of the elliptical shape and orientation used to define $pr$).  

\section{Chemical and radiative transfer modeling} \label{Sect:chemical_modeling}

We carried out chemical and radiative transfer modeling to reproduce the observed p$\rm NH_2D$ column density and line emission distribution. First, we produced simulated p$\rm NH_2D$ abundance distributions using the most recent version of our gas-grain chemical code \citep{Sipila20}, which includes an extensive description of deuterium and spin-state chemistry, of critical importance for the present work. Very briefly, we used the one-dimensional (1D) source model for L1544 presented by \citet{Keto2015}, which was divided into concentric shells; chemical simulations were carried out separately in each shell. This process yields time-dependent abundance profiles. The initial abundances used in the chemical modeling are displayed in Table~\ref{tab:initialabundances}. We assume a constant grain radius of 0.1\,$\mu$m, and a grain material density of 2.5\,g\,cm$^{-3}$. Other model details can be found in \citet{Sipila20} and are omitted here for brevity.

\begin{table}[t]
	\centering
	\caption{Initial fractional abundances (with respect to total H nuclei, $n_{\rm H}$).}
	\begin{tabular}{cc}
		\hline
		\hline
		Species & Abundance\\
		\hline
		$\rm H_2$ & $5.00\times10^{-1} \,^{(a)}$\\
		$\rm He$ & $9.00\times10^{-2}$\\
		$\rm C^+$ & $1.20\times10^{-4}$\\
		$\rm N$ & $7.60\times10^{-5}$\\
		$\rm O$ & $2.56\times10^{-4}$\\
		$\rm S^+$ & $8.00\times10^{-8}$\\
		$\rm Si^+$ & $8.00\times10^{-9}$\\
		$\rm Na^+$ & $2.00\times10^{-9}$\\
		$\rm Mg^+$ & $7.00\times10^{-9}$\\
		$\rm Fe^+$ & $3.00\times10^{-9}$\\
		$\rm P^+$ & $2.00\times10^{-10}$\\
		$\rm Cl^+$ & $1.00\times10^{-9}$\\
		\hline
	\end{tabular}
	\tablenotetext{a}{The initial $\rm H_2$ ortho/para ratio is $1 \times 10^{-3}$.}
	\label{tab:initialabundances}
\end{table}

\begin{figure*}
	\begin{minipage}[c]{0.42\textwidth}
	  \hspace*{-0.3cm} \includegraphics[width=1.1\textwidth]{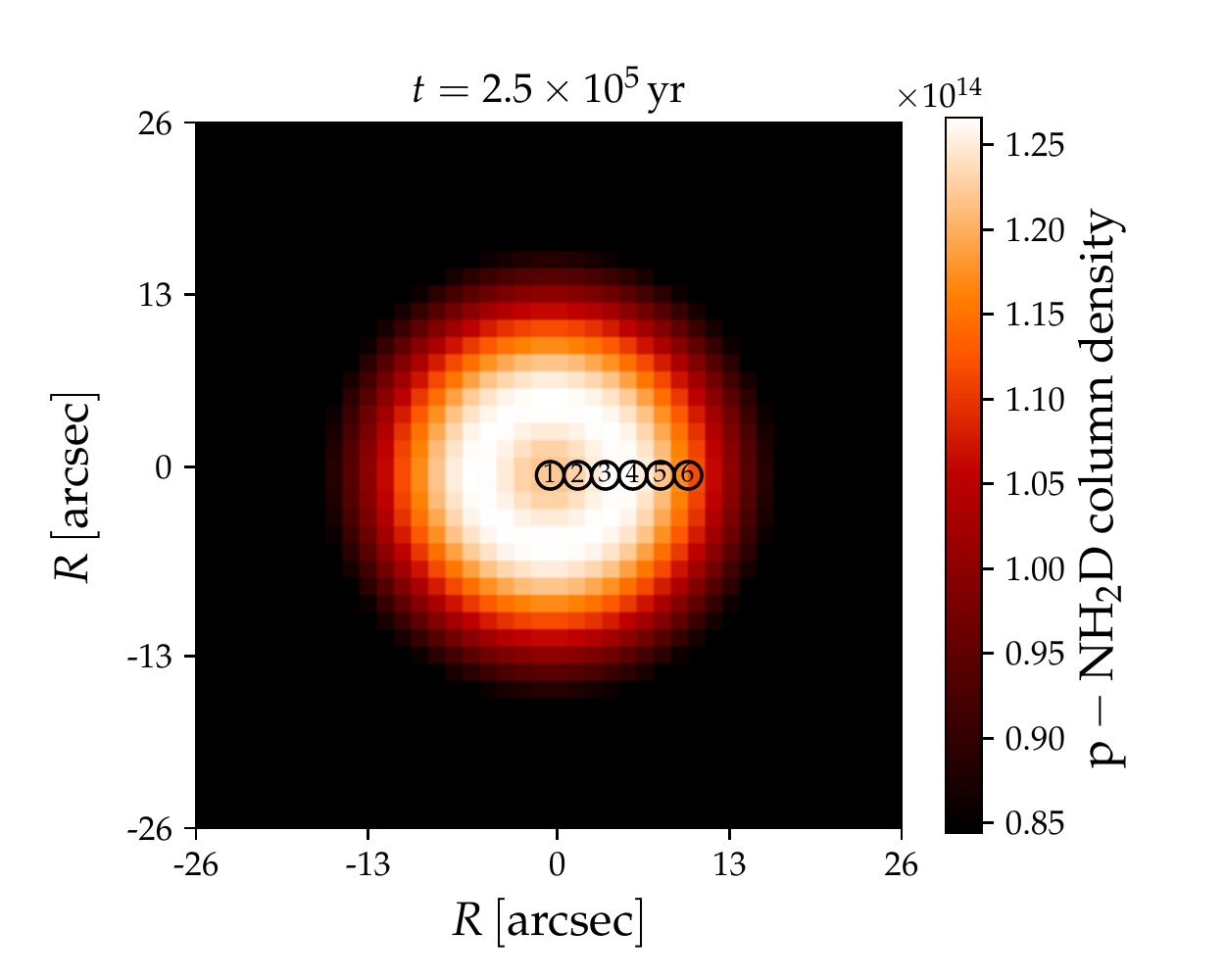}
	\end{minipage}
	\begin{minipage}[c]{0.58\textwidth}
	   \centering
	   \includegraphics[width=1\textwidth]{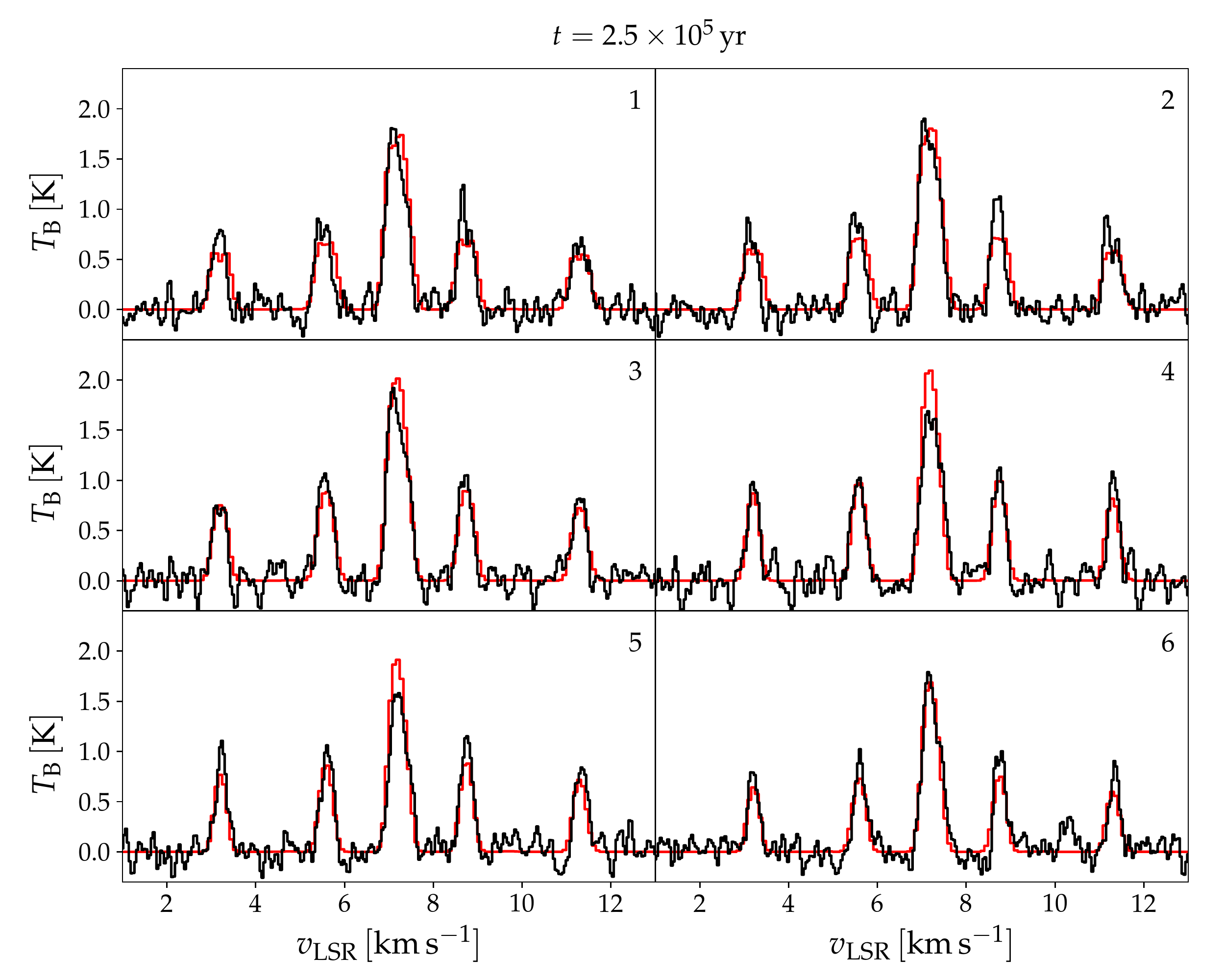}
	\end{minipage}
   \caption{{\sl Left:} Map of the p$\rm NH_2D$ column density distribution predicted by the chemical model at the best-fit time. The column densities have been convolved to a beam of 2.5$\arcsec$ (425\,au). The black circles with numbers are the areas within which the red spectra in the right panel, with the corresponding numbers, have been extracted. {\sl Right:} Simulated p$\rm NH_2D$ ($1_{11}-1_{01}$) line emission profiles (red) at $\sim$2.5$\arcsec$ intervals from the core center. The observed lines toward equivalent positions in the South-East direction in Figure\,\ref{fig:L1544_mosaic}, parallel to the major axis, are shown in black.}
    \label{fig:pNH2D_columnDensity}
\end{figure*}

We searched for the best fit to the observed line map by extracting the p$\rm NH_2D$ abundance profile from the chemical model at various time steps (between 10$^4$ and 10$^6$\,yr), and using the abundance data as input for the non-LTE line radiative transfer code LOC \citep{Juvela20} to simulate the line emission. We adopted the p$\rm NH_2D$ -- p$\rm H_2$ collisional coefficients from \citet{Daniel14} and assumed that the hyperfine components in each rotational level have the same excitation temperature. The hyperfine component frequencies and relative intensities have been adopted from \citet{Melosso2021-NH2D_frequencies}. The velocity profile of the contracting cloud (including a uniform level of turbulence of 0.1\,km\,s$^{-1}$) was taken from \citet{Keto2015}. With this setup, the best fit between the model and the observations is reached after a chemical evolution of $t \sim 2.5 \times 10^5 \, \rm yr$, as determined by a $\chi^2$ analysis applied to modeled and observed spectra toward six positions at and near the center of the core\footnote{The six observed spectra have been extracted from the six 2.5\arcsec \, circular areas, along the major axis and toward the South-East direction, displayed in Fig.\,\ref{fig:L1544_mosaic}. We also compared with the six spectra extracted from the perpendicular direction along the minor axis (see Fig.\,\ref{fig:L1544_mosaic}) and found similar results within the central 1800\,au of the kernel (see Appendix\,\ref{Appendix:full_mosaic} and Fig.\,\ref{fig:spectra_cut1}), while the outer positions show some deviation, suggesting that the (spherically symmetric) model is in better agreement with the physical structure along the major axis (see also Appendix\,\ref{Appendix:density_profile}).}. We stress that the chemical best-fit time of 2.5$\times$10$^5$\,yr cannot be used as a chemical clock or to deduce the dynamical age of the core. The reason for this is that the chemistry is run using a static physical structure as input \citep[a BE sphere with a density, temperature and velocity profile deduced by][]{Keto2015}, so chemical processes are not followed simultaneously with the dynamical evolution. Future work will focus on self-consistent chemical-dynamical evolution models, using hydro- and magneto-hydrodynamics simulations, of the types mentioned in Appendix \ref{Appendix:density_profile}.

\begin{figure}
    \centering
    \includegraphics[width=0.47\textwidth]{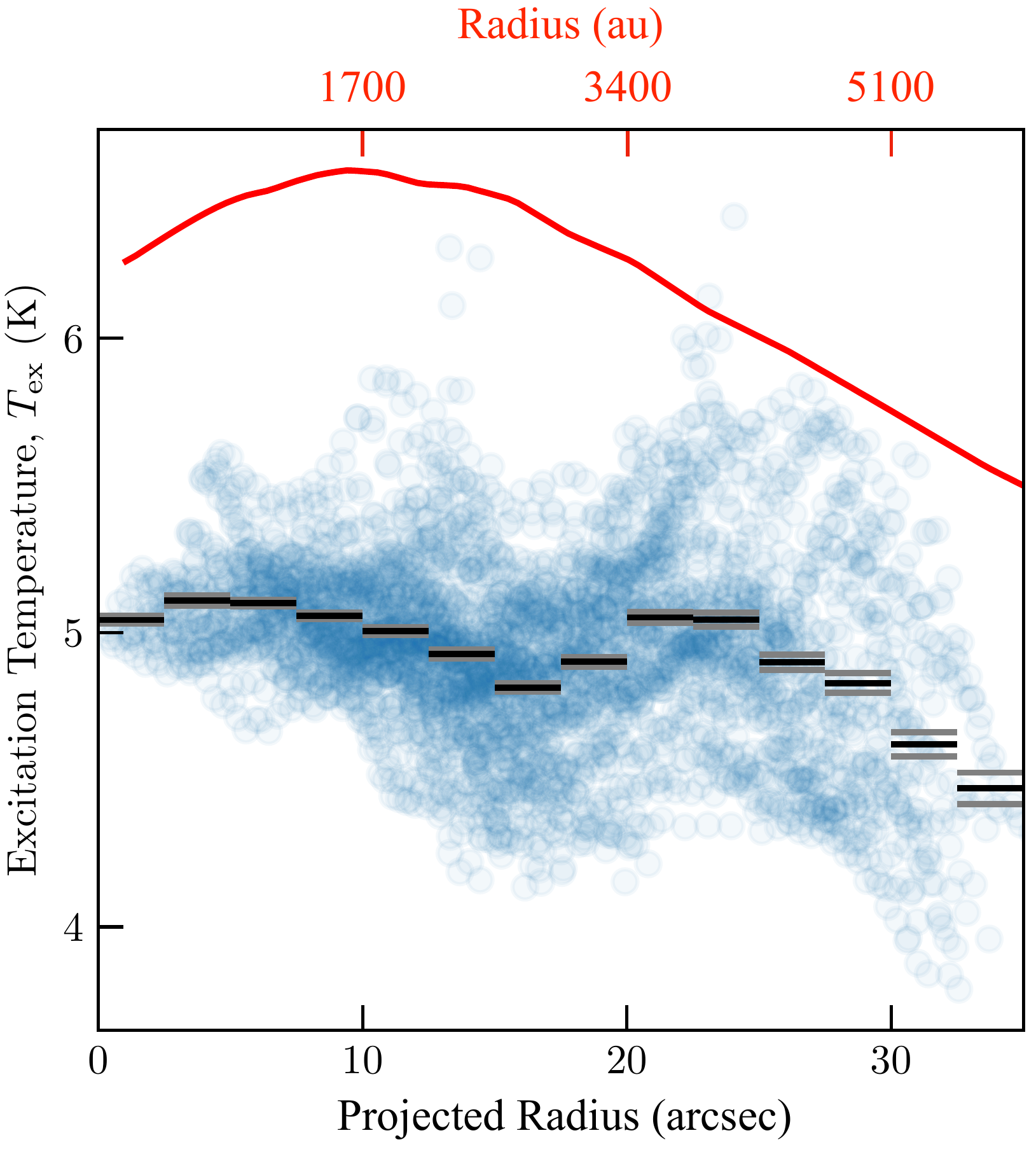}
    \caption{p\nhtdoo \, excitation temperature ($T_{ex}$) as a function of {\it projected radius} (see Section\,\ref{sec:linefit} for its definition). Blue circles correspond to individual measurements. The black and gray horizontal lines are the mean and corresponding uncertainty within a 2.5\arcsec \, bin. The red curve is the model $T_{ex}$ profile, the predicted excitation temperature as a function of core {\it radius}, $T_{ex}(r)$.}
    \label{fig:excitation_temperature}
\end{figure}

Figure~\ref{fig:pNH2D_columnDensity} shows the simulated p$\rm NH_2D$ column density map and the ($1_{11}-1_{01}$) line emission profiles at the best-fit time. 
The modeled  column densites and line emission maps agree well with the observations, showing a relatively flat column density map within the central $\simeq$13\arcsec \, (or about 2200\,au). In fact, although the pNH$_2$D column density map in Fig.\,\ref{fig:pNH2D_columnDensity}, left panel, shows a central valley within a bright ring of $\simeq$5\arcsec \, ($\simeq$850\,au), while our observations show a flattened structure (see Fig.\,\ref{fig:column_density_profile}), the model column density at the dust peak is only 4\% lower than at the bright ring (1.22$\times$10$^{14}$\,cm$^{-2}$ instead of 1.27$\times$10$^{14}$\,cm$^{-2}$), well within our calibration uncertainties. As we will show in Section\,\ref{Section:discussion}, this implies that our observations are consistent with $\rm NH_2D$ freeze-out in the central core. We have verified that the ($1_{11}-1_{01}$) line emission profiles cannot be reproduced by the models without $\rm NH_2D$ depletion toward the center (see Appendix\,\ref{Appendix:modeled_spectra}). We note that the observed average pNH$_2$D column density in Fig.\,\ref{fig:column_density_profile} is slightly (less than a factor of two) larger than that in Fig.\,\ref{fig:pNH2D_columnDensity} ($\simeq$2$\times$10$^{14}$ cm$^{-2}$ versus $\simeq$1.2$\times$10$^{14}$ cm$^{-2}$), which is once again underlying the good agreement between model and observations, despite the simplicity of the model. The measured and modeled excitation temperatures $T_{ex}$ are also similar, as can be seen in Fig.\,\ref{fig:excitation_temperature}, which compares the measured $T_{ex}$ (as deduced from the \verb+pyspeckit+ hfs fit of all the spectra within the map in Fig.\,\ref{fig:L1544_maps}) as a function of {\it projected radius} ($pr$, see Sect.\,\ref{sec:linefit}) with the excitation temperature profile {\it within} the core ($T_{ex}(r)$, with $r$ the core {\it radius}) predicted by the radiative transfer LOC. Although a one-to-one comparison between the two excitation temperatures cannot be made, as one should instead compare the measured $T_{ex}$ with the integral along the line of sight of the model $T_{ex}(b)$ for each impact parameter $b$, the two values are comparable, with the model $T_{ex}(r)$ within 20-30\% the measured $T_{ex}$.

A final note concerns the adopted physical structure. The \citet{Keto2015} 1D model used here provides self-consistently volume density and velocity profiles across the core and it can be easily input in comprehensive gas-grain chemical models such as the present one and previous work by, e.g., \citet{Vasyunin2017} on complex organic molecules in L1544. Naturally,  this 1D model cannot reproduce the elongated structure observed at 1.3mm \citep[see Figure 3 in][and Appendix\,\ref{Appendix:density_profile}]{Caselli2019-L1544_ALMA}. In fact, ALMA observations are better reproduced by 3D non-ideal magneto-hydrodynamic (niMHD) simulations of a contracting cloud with peak volume density of 10$^7$\,\cc \, (see their Figure 4). The 3D niMHD simulation gives rise to a pre-stellar core with an elongated flattened central structure, in harmony with the observed 1.3\,mm ALMA dust continuum emission. The adoption of this 3D model for the chemistry is however beyond the scope of this paper and it will be used in a future publication, where only simple species will be followed in a reduced chemical network to be included in the dynamical simulation. Despite the difference of the two (1D and 3D) models, it is interesting to note that the density and velocity profiles along the major and minor axis of the simulated 3D core do not differ substantially from those predicted by the 1D model of \citet{Keto2015}, as shown in Fig.\,\ref{fig:L1544_profiles}. In particular, the density and velocity profiles along the major axis of the 3D model are especially close to our adopted 1D model (see Fig.\,\ref{fig:L1544_profiles}), suggesting  similarities that, together with the radiative transfer results shown in Fig.\,\ref{fig:pNH2D_columnDensity}, make our 1D model a very good approximation of the L1544 structure.   

\section{Discussion: almost complete freeze-out in the kernel}
\label{Section:discussion}

In the previous sections we showed that ALMA has allowed us to resolve for the first time the distribution of deuterated ammonia in the central 1000\,au of the pre-stellar core L1544. Despite the complex morphology of the L1544 core as viewed at the frequency of the pNH$_2$D ($1_{11}-1_{01}$) line (see Fig.\,\ref{fig:L1544_maps}), it is clear that the pNH$_2$D column density has a flattened distribution in correspondence of the dust continuum peak, where the kernel is located (see also Fig.\,\ref{fig:column_density_profile}). The radius of the kernel is about $\simeq$1800\,au, which is $\sim$5 times smaller than the CO depletion zone observed toward the same pre-stellar core by \citet{CaselliWalmsley1999}. This is the reason why previous interferometric observations from \cite{Crapsi2007-L1544}, done with poorer angular resolution (5.8\,\arcsec $\times$4.5\,\arcsec), could not reveal the abundance drop of deuterated ammonia, as demonstrated in Appendix\,\ref{Appendix:model_fivearc}. 

In the following, we will present the fractional abundance profiles predicted by our chemical model, at 2.5$\times$10$^5$\,yr, of all ammonia isotopologues as well as the profile of the {\it total depletion factor}, $f_{\rm D_{tot}}$, defined by: 
\begin{equation}
    f_{\rm D_{tot}} = \frac{\sum_i [x(i)_{\rm gas} + x(i)_{\rm solid}]}{\sum_i x(i)_{\rm gas}},
\end{equation}
where $i$ represents any species in the chemical model containing an element heavier than He. Thus, $f_{\rm D_{tot}}$ quantifies the amount of elements heavier than He subsisting in the gas phase, with the higher values corresponding to higher levels of depletion. 

\begin{figure*}
    \centering
    \includegraphics[width=0.95\textwidth]{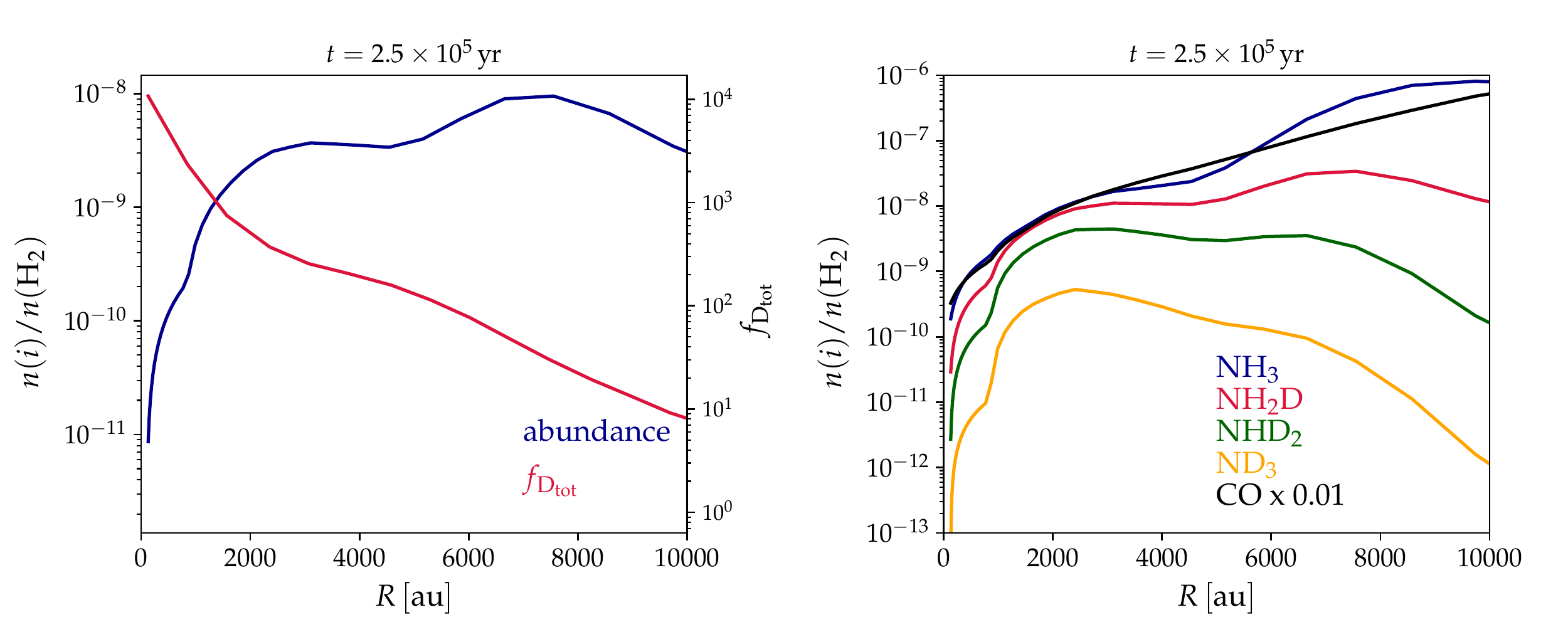}
    \caption{{\sl Left}: Model fractional abundance of p$\rm NH_2D$ (blue) and overall gas-phase depletion factor (red) at the best-fit time as a function of radius. {\sl Right}: Fractional abundances w.r.t. H$_2$ of normal and deuterated ammonia (colors, indicated in the Figure), as well as of CO (multiplied by 0.01, black), at the best-fit time as a function of distance from the core center. The ammonia fractional abundances represent sums over the nuclear spin forms (ortho and para; also meta for $\rm ND_3$).}
    \label{fig:L1544_abu+fD}
\end{figure*}

 The abundance profile of p$\rm NH_2D$ at the best-fit time is displayed in Fig.\,\ref{fig:L1544_abu+fD} along with $f_{\rm D_{tot}}$. It is interesting to note that the pNH$_2$D abundance starts to drop already within about 7000\,au, whereas the integrated intensity map as well as the simulated observations show a centrally concentrated structure up to the central $\simeq$2000\,au. The abundance drop is in fact partially hidden to observations due to the radiative transfer of the p\nhtdoo \ line. In particular, despite a ``catastrophic" pNH$_2$D freeze-out present within the central 2000\,au (see Fig.\,\ref{fig:L1544_abu+fD}), the corresponding decrease in column density is only $\sim$4\%; this is shown in the left panel of Fig.\,\ref{fig:pNH2D_columnDensity}, where a relatively bright ring of about 5\arcsec \, in radius, within the pNH$_2$D depletion zone, is clearly visible. This is due to the fact that the column density is dominated by larger scales, where p$\rm NH_2D$ is abundant. The critical density of the ($1_{11}-1_{01}$) line is $\sim1.4 \times 10^5 \, \rm cm^{-3}$ at $T = 10\,\rm K$, and so collisional excitation dominates in a region roughly 13\arcsec \ in radius, enclosing the bright ring in the left panel of Fig.\,\ref{fig:pNH2D_columnDensity}. The inferred column density ring and shallow inner valley (Fig.\,\ref{fig:pNH2D_columnDensity}) are due to the drop in the pNH$_2$D abundance toward the center of the core. The simulated main hyperfine component peak has an optical depth close to 1, in agreement with our observations, and the strength of the whole line emission  follows the morphology of the column density distribution. 
 
 The abundance distributions of the modeled normal and deuterated ammonia (summed over the spin states) as well as that of CO are also shown in Fig.\,\ref{fig:L1544_abu+fD} (right panel); it is evident that the depletion of singly-deuterated ammonia within the central 2000\,au is not due to conversion into multiply-deuterated ammonia, but is instead a result of freeze-out onto grain surfaces. These results, including the ortho-to-para ratios of NH$_2$D, are consistent with those obtained by \citet{Hily-Blant2018}, who modelled the molecular composition and the nuclear spin chemistry of a collapsing Bonnor-Ebert \citep[BE;][]{Bonnor1956,Ebert1955} sphere.
  According to our model, the total depletion factor $f_{\rm D_{tot}}$ increases from 1000 to 10000 in the central 2000\,au, implying that between 99.9\% and 99.99\% of all the species heavier than He are predicted to reside on the surface of dust grains within the kernel. This has the consequence that thick icy mantles are expected to form on top of dust grains within the central regions of pre-stellar cores, just before the formation of a protostar and protostellar disk. Thick icy mantles could be responsible for the change in dust opacity measured toward the center of L1544 by \citet{Chacon2019}. 

We note that, previous to this work, coarser interferometric observations of pNH$_2$D showed a clear centrally peaked distribution toward L1544 \citep{Crapsi2007-L1544}. For this reason, pNH$_2$D was considered the species most resilient to freeze-out \citep[compared with all other species observed so far toward this core which all show direct or indirect evidence of depletion, including N$_2$D$^+$; e.g.][]{SpezzanoCaselli2017,Redaelli2019}. Now that even p\nhtd \, freeze-out has been unveiled and our modeling reproduces the observed column density map and line profiles of p\nhtdoo , we can conclude that the almost complete freeze-out predicted by our chemical model is then consistent with the data available for this source. Moreover, as already mentioned, previous single-dish and interferometric observations suggested a NH$_3$ abundance profile increasing toward the central regions \citep[e.g.,][]{Crapsi2007-L1544}. To investigate this apparently contradicting result, we simulate also the NH$_3$(1,1) line profiles using the pNH$_3$ abundance profile predicted by our model, after 2.5$\times$10$^{5}$\,yr of chemical evolution, and compare them with the VLA-only observations in Figure 2 of \cite{Crapsi2007-L1544}. We also consider the case without NH$_3$ freeze-out in the central 6000\,au (which differs from our predicted profile shown in Fig.\,\ref{fig:L1544_abu+fD}, right panel) and see if this better reproduces the observations, in line with the conclusions from \cite{Crapsi2007-L1544}. The assumed beam for the simulated observations of NH$_3$(1,1) is 4\arcsec , close to the VLA resolution of 4.35\arcsec $\times$ 3.45\arcsec. The spectra are reported in Fig.\,\ref{fig:nh3_spectra}, showing that our model prediction of NH$_3$ freeze-out toward the center (see Fig.\,\ref{fig:L1544_abu+fD}) is consistent with the \citet{Crapsi2007-L1544} observations, including the relative intensity of the groups of hyperfines and the negligible difference between the spectrum centered at the dust peak position and the one at the 22.3\arcsec \, offset. The main difference is that our simulated lines are about two times brighter than the observed ones, but this is expected due to the fact that we are not simulating interferometric observations, and the core envelope is significantly contributing to the simulated line flux (unlike in the case of interferometric observations). In fact, higher sensitivity VLA data of NH$_3$(1,1) toward L1544, combined with GBT single-dish data, show a peak temperature in better agreement with our modeling; these data will be presented in an upcoming paper focusing on a new gas temperature map of L1544 (Schmiedeke et al., in prep.). Another interesting result of this comparison is that variations in the abundance profile of ammonia in the central 5000\,au do not change the spectra (see dashed histograms in Fig.\,\ref{fig:nh3_spectra}, which refer to a model with a flat fractional abundance of NH$_3$ of 2$\times$10$^{-8}$ within the central 5000\,au), demonstrating that NH$_3$(1,1) cannot be used to measure its abundance profile within the central few thousand au, because of its relatively low critical density (see Appendix \ref{Appendix:nh3_spectra}). Interestingly, new high-resolution and high-sensitivity NH$_3$ observations of the pre-stellar core H-MM1 in Ophiuchus show a clear depletion zone towards its dust peak (Pineda et al., in prep.), thus confirming our predictions.

A final comment should be added about the velocity structure used as input to the chemical model. In some previous studies of L1544, the \citet{Keto10a} infall velocity profile had to be scaled up by a factor of $\sim$2 to reproduce observations obtained with single-dish telescopes \citep{Bizzocchi13,Redaelli2019}. In the present case, a similar scaling leads to double-peaked line profiles, even for the satellite components, that are completely excluded by the observations. This discrepancy with the earlier single-dish studies is likely related to the 1D spherically symmetric structure of the source model (while the source itself is certainly not spherical on the large scales traced by single-dish observations), and to the possible presence of more complex velocity fields at the core scale (see Appendix\,\ref{Appendix:density_profile}). The present interferometric observations probe a relatively small volume, where the number density is close to the critical density of the observed transition (1.4$\times$10$^5$\,cm$^{-3}$), i.e. a region with radius of about 4000\,au, considering the adopted density profile shown in Fig.\,\ref{fig:L1544_profiles} (red-dashed curve); hence the 1D symmetry assumption is adequate, as already discussed. Therefore, the \citet{Keto10a} physical model, which already successfully reproduced line profiles of several species observed with single-dish telescopes toward L1544 \citep[e.g.,][]{Caselli2012,Keto2015}, still remains valid for these new ALMA interferometric observations, as well as recent work from \citet{Redaelli2021}.

\section{Conclusions}

Our ALMA spectroscopic observations, coupled with chemical and radiative transfer modeling, of the prototypical pre-stellar core L1544 have unveiled the depletion zone of deuterated ammonia, with size very close to the kernel previously found with ALMA continuum observations (radius $\simeq$1800\,au). This has not been possible before because of the poorer angular resolution (factor of $\geq$2) and quality of previous observations of the same line. Our chemical model, applied to the physical structure of L1544, reproduces remarkably well the p\nhtdoo \, emission. Simulated observations of the modelled pre-stellar core show that the high p\nhtd \, fractional abundance in regions outside the kernel, with volume densities close to the critical density of the (1$_{11}$-1$_{01}$) transition, prevent a clear view of the actual catastrophic \nhtd \, freeze-out region without high enough angular resolution. We find that more than 99.9\% of all species heavier than He reside in thick icy mantles within the central kernel, the future stellar cradle. In pre-stellar cores like L1544, with central volume densities $\geq$10${^6}$\,cm$^{-3}$ and corresponding freeze-out time scales $\leq$10$^{3}$\,yr, an almost complete freeze-out is therefore inevitable, just before star formation. The thick icy mantles \cite[about 150 monolayers on the 0.1\,$\mu$m-grains in our chemical model, close to the predictions from][]{Draine1985} may promote coagulation of dust grains \citep[e.g.,][]{Ormel2009,Chacon2019} and allow preservation and delivery of the frozen pre-stellar chemistry into the next stages of evolution, when the protostar and protoplanetary disk will form. Some of this pre-stellar ice, especially that trapped within icy pebbles in the outer part of the disk, may survive later stages of planet formation and evolution. Indeed, similarities in molecular composition and isotopic fractionation between  star- and planet-forming regions and primitive material in our Solar System, such as comets and carbonaceous chondrites  
\citep[e.g.][]{Mumma2011,CaselliCeccarelli2012,Ceccarelli2014,vanDishoeck2014,Oberg2015,Altwegg2019,Cleeves2014,Drozdovskaya2021}, strongly hint at a crucial role of pre-stellar chemistry in shaping the composition and further chemical/minerology evolution of the building blocks of planets.

\acknowledgments
We are grateful to Malcolm Walmsley, for the many inspiring discussions on complete depletion that lead to this work. We thank the anonymous referee and Doug Johnstone for a final review which improved the clarity of the paper. We gratefully acknowledge the support of the Max Planck Society. I.J.-S. has received partial support from the Spanish State Research Agency (PID2019-105552RB-C41). ZYL is supported in part by NASA 80NSSC18K1095 and NSF AST-1910106. This paper makes use of the following ALMA data: ADS/JAO.ALMA\#2013.1.01195.S and ADS/JAO.ALMA\#2016.1.00240.S. ALMA is a partnership of ESO (representing its member states), NSF (USA) and NINS (Japan), together with NRC (Canada), MOST and ASIAA (Taiwan), and KASI (Republic of Korea), in cooperation with the Republic of Chile. The Joint ALMA Observatory is operated by ESO, AUI/NRAO and NAOJ.

\bibliography{bibliography}{}
\bibliographystyle{aasjournal}

\appendix

\section{Full mosaic map} \label{Appendix:full_mosaic}
\begin{figure}
    \centering
    \includegraphics[width=0.8\textwidth]{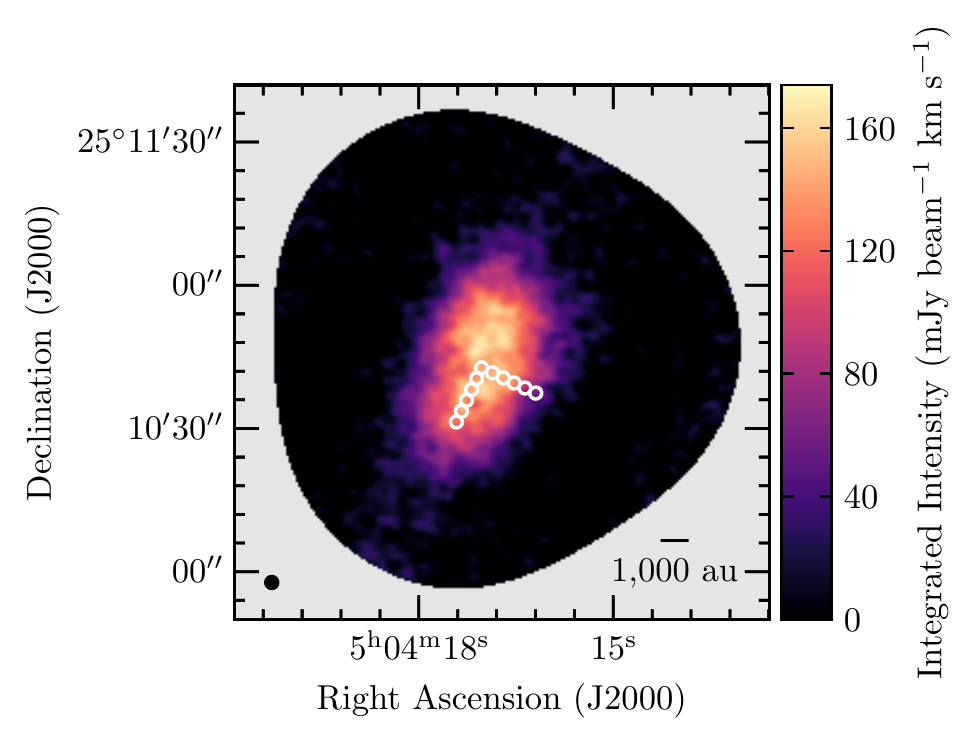}
    \caption{Integrated intensity map of the pNH$_2$D (1$_{11}$-1$_{01}$) transition on L1544, showing the 3-point mosaic. The white circles overlaid to the integrated intensity map mark the positions where the spectra of Fig.\,\ref{fig:pNH2D_columnDensity} (South-East direction, along the major axis) and Fig.\,\ref{fig:spectra_cut1} (South-West direction, along the minor axis) have been extracted.}
    \label{fig:L1544_mosaic}
\end{figure}

\begin{figure}
    \centering
    \includegraphics[width=0.7\textwidth]{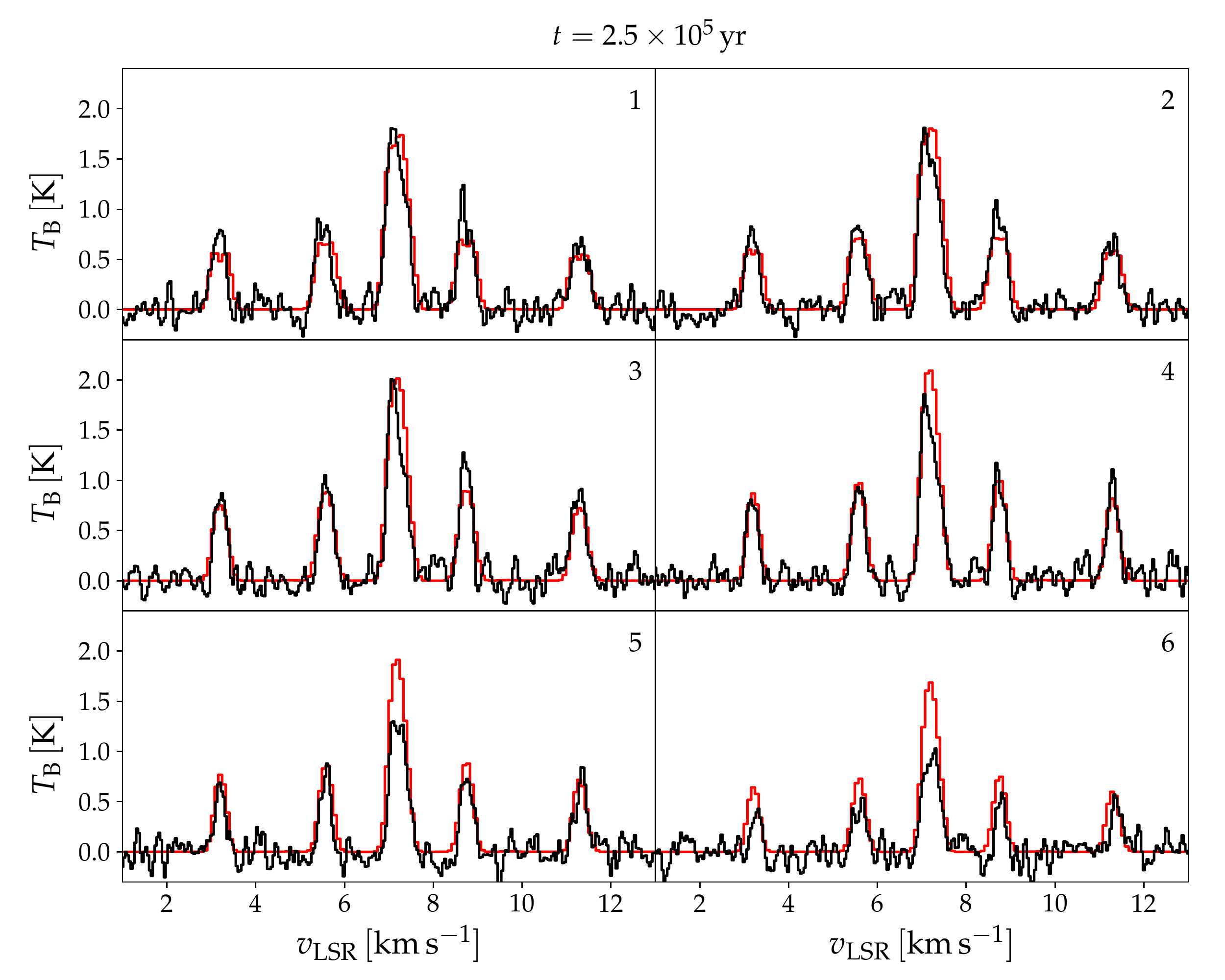}
    \caption{Same as the right panel of Fig.\,\ref{fig:pNH2D_columnDensity} but now the black histograms are the spectra extracted from the six circles in the South-West direction in Fig.\,\ref{fig:L1544_mosaic} (along the L1544 minor axis). Note that the observed spectra in the outer positions (numbers 5 and 6) are now less intense than those predicted by our model. This suggests that the major axis of L1544 has physical and chemical structure in better agreement with our 1D model.}
    \label{fig:spectra_cut1}
\end{figure}

\restartappendixnumbering
In Figure \ref{fig:L1544_maps}, a zoom-in view of the p\nhtdoo \ integrated intensity map is shown, in order to highlight the structure and better compare with the column density map. In this Appendix, we present the full 3-point mosaic of the same map to show the extent of the mapped area (see Fig.\,\ref{fig:L1544_mosaic}). The six white circles along the major axis in the South-East direction in Fig.\,\ref{fig:L1544_mosaic} enclose the areas where the spectra have been extracted for comparison with the model spectra (see Fig.\,\ref{fig:pNH2D_columnDensity}). The spectra from the other six white circles, those along the minor axis, are reported in Fig.\,\ref{fig:spectra_cut1} and compared with the same model spectra as in Fig.\,\ref{fig:pNH2D_columnDensity}. It is apparent that the outer positions in Fig.\,\ref{fig:spectra_cut1} (numbers 5 and 6) show model spectra brighter than the observed ones, suggesting that the major axis has a physical and chemical structure closer to our spherically symmetric model. This is supported also by Fig.\,\ref{fig:L1544_profiles}, where the density and velocity profiles of the 3D model are compared with our 1D model profiles. 

\restartappendixnumbering

\section{Comparison with single dish \nhtdoo \, spectrum} \label{Appendix:single_dish}

\begin{figure}
    \centering
    \includegraphics[width=0.7\textwidth]{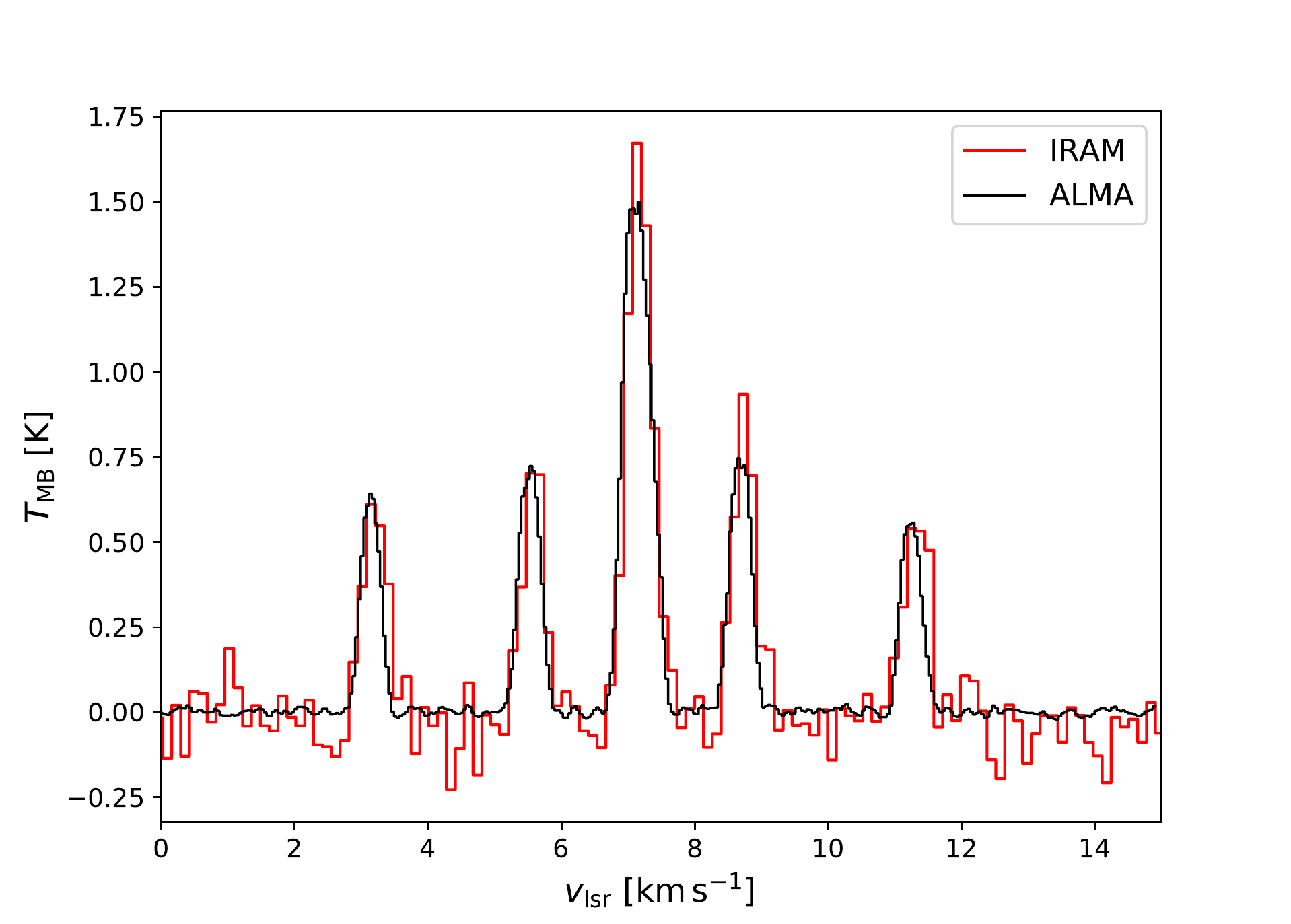}
    \caption{Comparison between the p\nhtdoo \, IRAM-30m spectrum (red histogram) and the ALMA 12\,m+ACA
spectrum within the same 22\arcsec \, beam of the 30m observations (black histogram), obtained toward the L1544 kernel. This clearly shows that the ALMA 12\,m+ACA combined mosaic fully recovers the line flux.}
    \label{fig:L1544_spectrum}
\end{figure}

Figure \ref{fig:L1544_spectrum} shows the comparison between the p\nhtdoo \, spectrum obtained with the IRAM-30m single dish observations (with 22\arcsec \, beam) and the one obtained by ALMA (12\,m+ACA) within the same area. The IRAM 30m spectrum of p\nhtdoo \ has been extracted at the dust peak of an On-The-Fly map covering the inner 22\arcsec $\times$ 22\arcsec \ of L1544 around the dust peak. The emission map was obtained in the framework of the project 013-13 (PI: S. Spezzano) in October 2013. Position switching was used, with the reference position set at (-180\arcsec, 180\arcsec) offset with respect to the map centre. The EMIR E090 receiver was used with the Fourier Transform Spectrometer backend (FTS) with a spectral resolution of 50 kHz. The mapping was carried out in good weather conditions ($\tau \sim$ 0.03) and a typical system temperature of $\sim$90-100 K.  It is clear from the figure that all the flux is fully recovered, demonstrating that ALMA mosaicing, together with the combination of 12\,m and ACA data, provides a powerful technique to study pre-stellar cores in detail. 

\restartappendixnumbering

\section{Modeled spectra in case of no central freeze-out}
\label{Appendix:modeled_spectra}

\begin{figure}
    \centering
    \includegraphics[width=0.7\textwidth,angle=0]{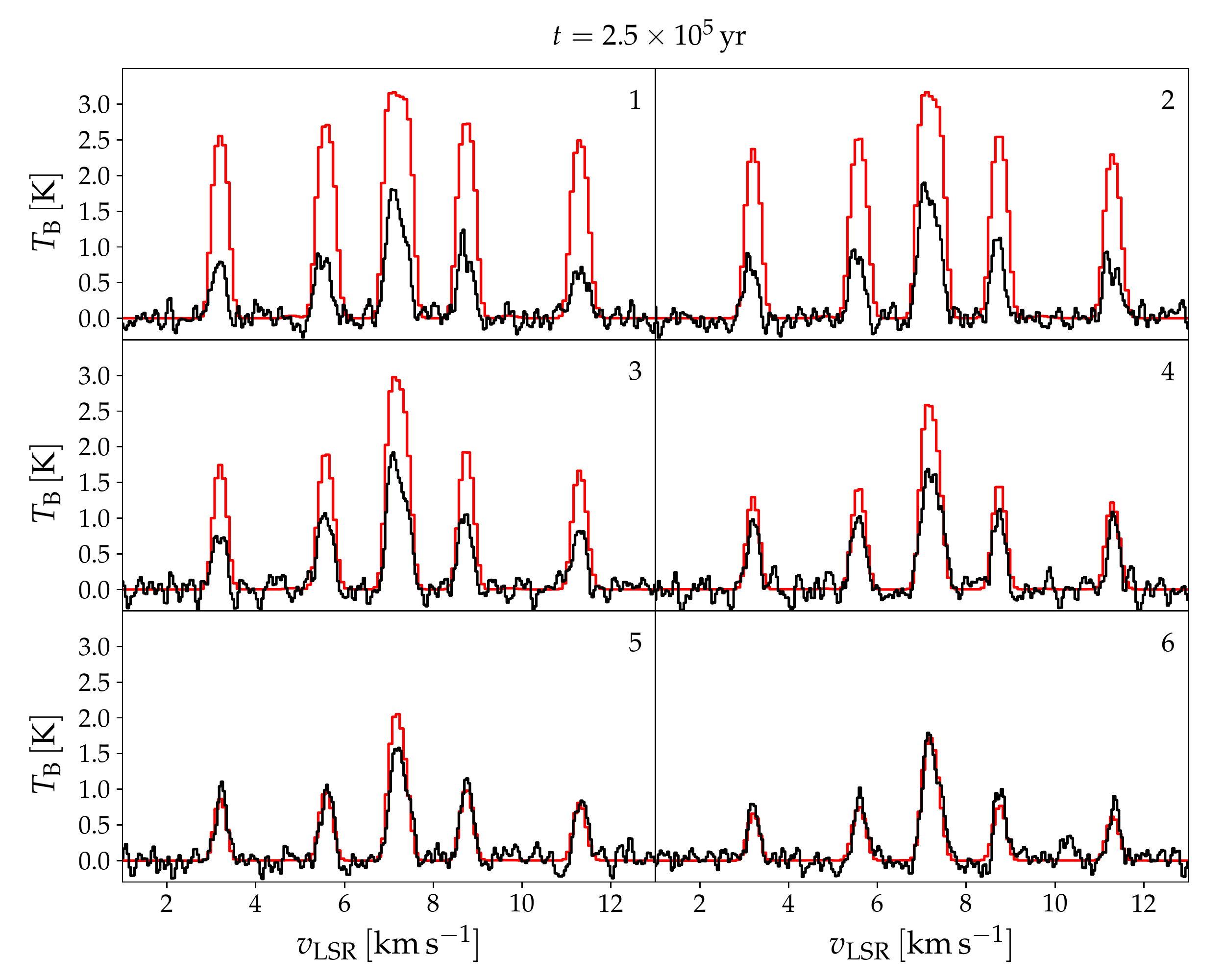}
    \caption{Same as Fig.\,\ref{fig:pNH2D_columnDensity}, right panels, but assuming that the abundance of p$\rm NH_2D$ is constant within the central 2000\,au, instead of dropping as predicted by our chemical models (see Fig.\,\ref{fig:L1544_abu+fD}). The modeled lines clearly overestimate the observed spectra in the central four positions, demonstrating that freeze-out is needed.}
    \label{fig:spectra_nofreezeout}
\end{figure}

As shown in Fig.\,\ref{fig:pNH2D_columnDensity}, our model reproduces very well the p\nhtdoo \ spectra toward the L1544 kernel. This model predicts that \nhtd \ (as well as other species containing elements heavier than He) quickly disappears from the gas phase within the central 2000\,au because of freeze-out onto dust grains (see Fig.\,\ref{fig:L1544_abu+fD}). To test that central freeze-out is indeed needed to reproduce observations, Figure~\ref{fig:spectra_nofreezeout} shows the predicted p\nhtdoo \ spectra when the \nhtd \  fractional abundance is kept at the constant value of 6$\times$10$^{-9}$ within the central 2000\,au, instead of dropping as in Fig.\,\ref{fig:L1544_abu+fD}. The predicted spectra clearly overestimate the observed intensities in the central four positions, thus demonstrating that observations can only be reproduced when freeze-out is taken into account, as predicted by our chemical models. 

\restartappendixnumbering

\section{Density and velocity profiles} \label{Appendix:density_profile}

The physical structure of the L1544 core used for the chemistry code has been described in Sect.\,\ref{Sect:chemical_modeling}. As discussed there, the density profile from the self-consistent hydro-dynamic simulation of a contracting Bonnor-Ebert \citep[BE;][]{Bonnor1956,Ebert1955} sphere in quasi-static equilibrium \citep{Keto10a} well reproduces line profiles previously observed toward the L1544 dust peak \citep[e.g.][]{Keto10a,Caselli2012,Keto2015}. However, because of the spherical symmetry of the BE model, the elongated 1.3\,mm dust continuum emission observed at high angular resolution with ALMA cannot be reproduced \citep{Caselli2019-L1544_ALMA}. Indeed, \citet{Caselli2019-L1544_ALMA} showed that the central kernel is consistent with flattening produced by non-ideal magneto-hydrodynamic (niMHD) simulations of the contraction of a rotating magnetized cloud core. The peak density within the flattened structure in the 3D niMHD simulation reaches 10$^7$\,\cc, in agreement with the peak density at the center of the BE sphere, while the average density within the kernel is about 10$^6$\,\cc, consistent with previous single-dish observations \citep{Crapsi2007-L1544,Chacon2019}, which did not have high enough angular resolution to detect the central density enhancement. 

\begin{figure}
    \centering
    \includegraphics[width=0.7\textwidth,angle=270]{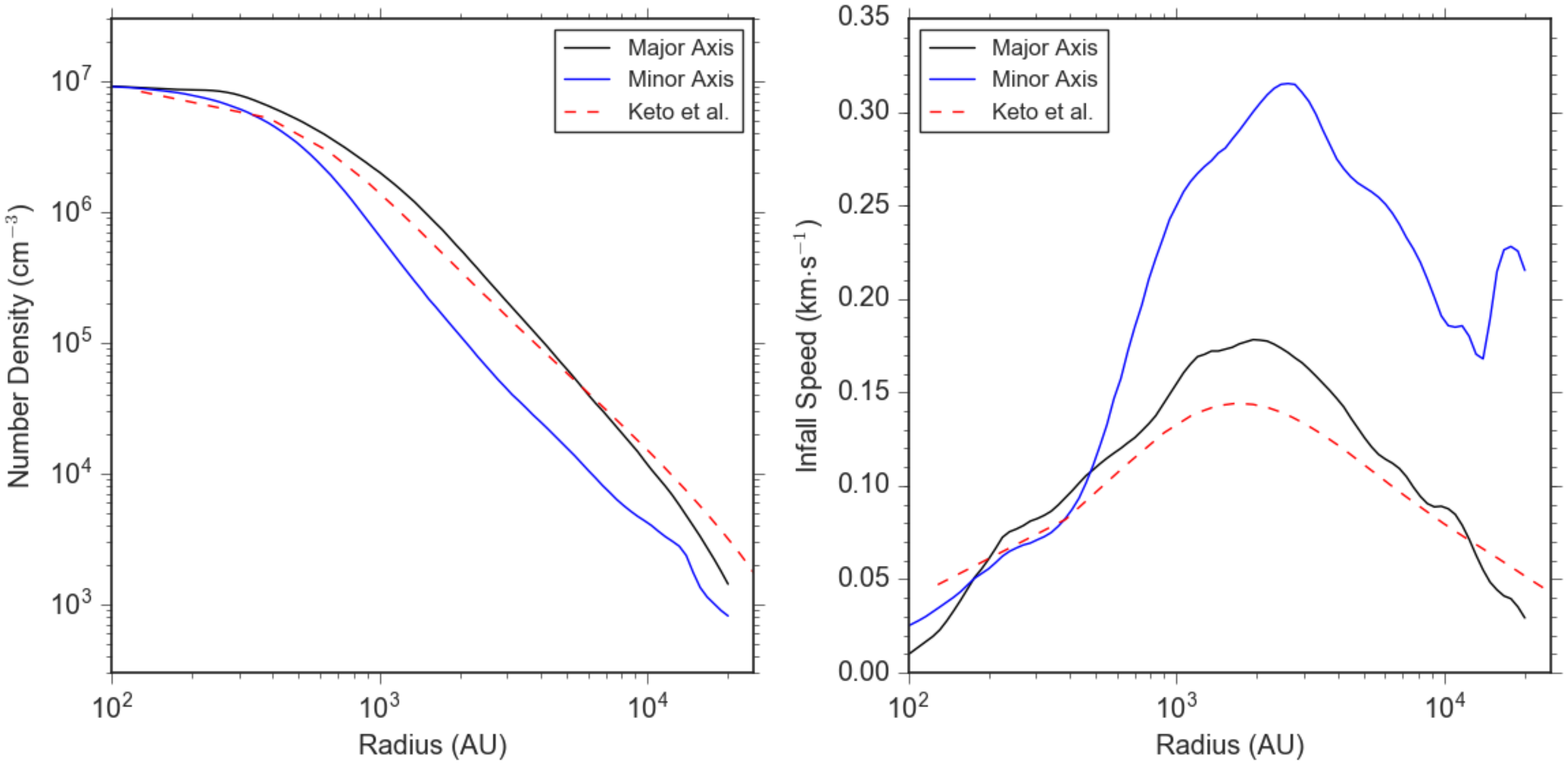}
    \caption{Comparison between the H$_2$ number density ({\it left panel}) and velocity ({\it right panel}) radial profiles of the BE sphere in quasi-static contraction from the HD simulations of \citet{Keto2015}, after 6$\times$10$^{5}$\,yr of evolution of an initial BE sphere with central density 1$\times$10$^4$\,cm$^{-3}$, radius 2$\times$10$^5$\,au and total mass 10\,M$_{\odot}$ (dashed red curves), and those along the major (black curves) and minor (blue curves) axis of the niMHD simulation of a magnetised contracting core, after 1.4$\times$10$^6$\,yr of evolution starting from a uniform 8.1\,M$_{\odot}$ cloud with volume density 2$\times$10$^4$\,cm$^{-3}$, radius 5$\times$10$^4$\,au and initial mass-to-flux ratio of 1, which best fits the ALMA 1.3\,mm dust continuum emission \citep[see][for details]{Caselli2019-L1544_ALMA}. Note the overall similarities despite the very different simulations (HD vs. niMHD) and different initial conditions.}
    \label{fig:L1544_profiles}
\end{figure}

Despite the structural difference between the central 1000\,au of the \citet{Keto2015} model (used as input of our chemical model) compared with the 3D niMHD simulated core (which better reproduces the ALMA 1.3\,mm dust continuum emission), the agreement between model and observed spectra in Fig.\,\ref{fig:pNH2D_columnDensity} is excellent. This is partly due to the choice of the cut, taken along the (projected) major axis of the kernel, as this is the direction where the density profile of the 3D structure more closely resembles the BE sphere (see Fig.\,\ref{fig:L1544_profiles}, left panel). The agreement with observed spectra along the minor axis becomes poorer (than that with observed spectra along the major axis) in the outer two positions, as seen in Fig.\,\ref{fig:spectra_cut1}. To reproduce in detail the spectra along the minor axis, we will have to include the chemistry in the 3D niMHD simulation, and this will be done in a future paper. For now, it is interesting to point out that the BE sphere in quasi-static contraction of \cite{Keto2015} has a density profile just in between the minor and major axis of the 3D niMHD simulated core (Fig.\,\ref{fig:L1544_profiles}, left panel). Also the radial velocity profile of the contracting BE sphere is very similar to that along the minor and major axis within a radius of 500\,au  (Fig.\,\ref{fig:L1544_profiles}, right panel); above this radius, the velocity profile of the contracting BE sphere follows more closely that along the major axis of the 3D niMHD core and it is only a factor of 2 away from that along the minor axis.  Considering the simplicity of the contracting BE sphere HD simulation compared with the niMHD simulation of the contracting magnetized core, it is striking to see how closely the two resemble each other. 

\section{Model predictions assuming a 5\arcsec $\times$ 5\arcsec \, beam}
\label{Appendix:model_fivearc}
\begin{figure}
    \centering
    \includegraphics[width=0.9\textwidth,angle=0]{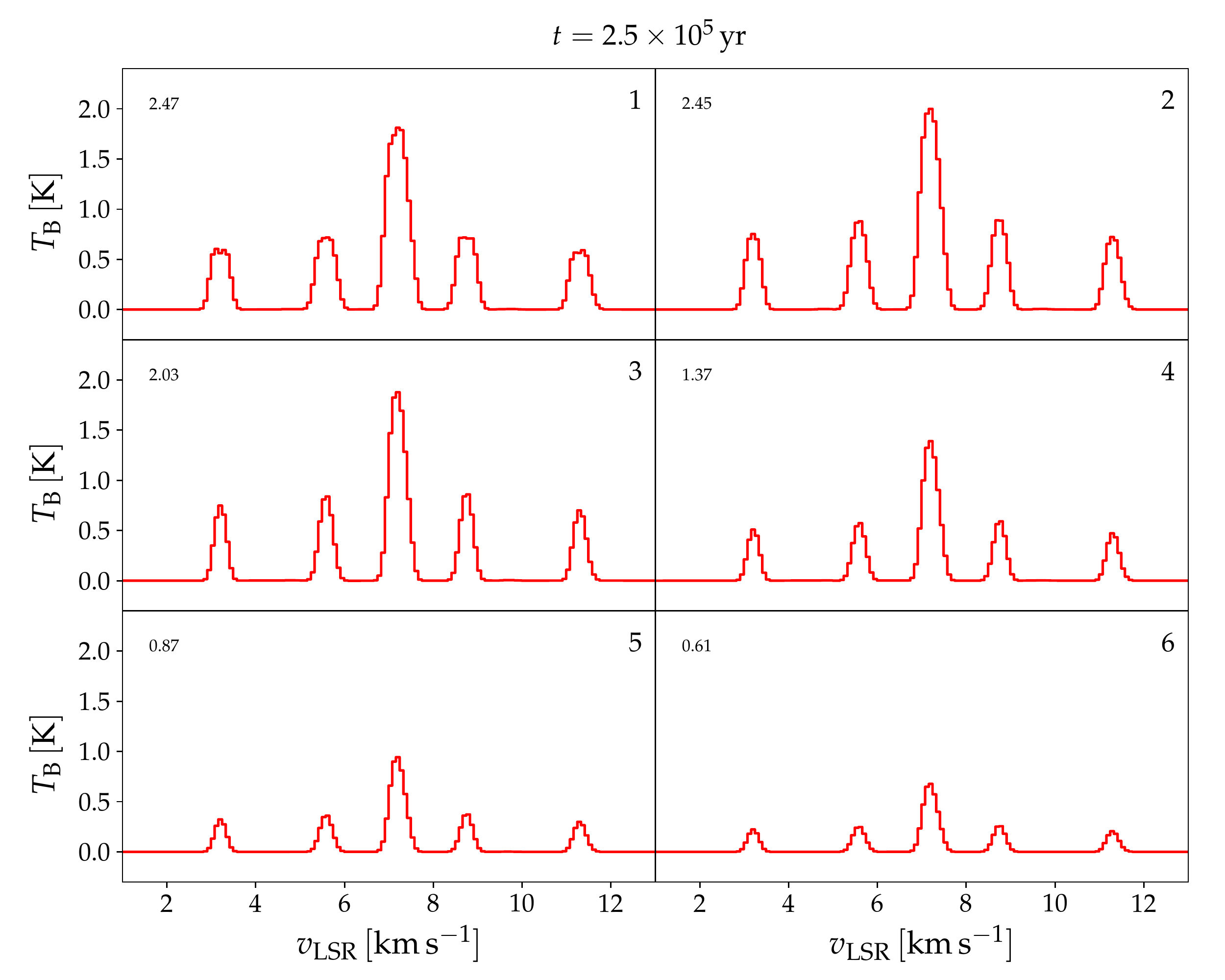}
    \caption{Simulated spectra of the same L1544 core model presented in Section\,\ref{Sect:chemical_modeling} and Fig.\,\ref{fig:pNH2D_columnDensity} but now assuming a telescope beam of 5\arcsec $\times$ 5\arcsec , similar to the one used by \citet{Crapsi2007-L1544}. The number in the top-left of each panel is the integrated intensity in K\,km\,s$^{-1}$, while the number in the top right shows the corresponding position across the strip labelled in Fig.\,\ref{fig:pNH2D_columnDensity}, left panel, but with a spacing of 5\arcsec \, instead of 2.5\arcsec . Although the simulated spectra look similar to those in Fig.\,\ref{fig:pNH2D_columnDensity}, the difference in integrated intensity between the spectrum extracted toward the core center (position 1) and the one 5\arcsec \, away (position 2, the integrated intensity peak) is very small. Thus, the flattening zone cannot be detected with the $\sim$5\arcsec \, beam of the \citet{Crapsi2007-L1544} observations.} 
    \label{fig:fivearc}
\end{figure}

\restartappendixnumbering
\citet{Crapsi2007-L1544} observed the same transition of deuterated ammonia using the IRAM Plateau de Bure interferometer with an angular resolution of 5.8\arcsec $\times$4.5\arcsec . To demonstrate that this is not sufficient to detect the pNH$_2$D depletion zone, we perform simulated observations of the same modelled core described in Sect.\,\ref{Sect:chemical_modeling}, assuming a spherical half power beam width of 5\arcsec. Fig.\,\ref{fig:fivearc} presents six simulated spectra extracted from six positions similar to those shown in Fig.\,\ref{fig:pNH2D_columnDensity}, but now separated by 5\arcsec \, (instead 2.5\arcsec). A 5\arcsec \, beam does not have a substantial effect on the simulated lines toward the center of the core (when compared with the red spectra in Fig.\,\ref{fig:pNH2D_columnDensity}). The most noticeable difference is that the satellite components become slightly less bright (and less optically thick) compared to the simulations with a 2.5\arcsec \, beam; the line brightness now peaks toward the central two positions (number 1 and 2 in Fig.\,\ref{fig:fivearc}). An integrated intensity map would then show a peak at the center and a monotonic drop toward the outer parts, consistent with the \citet{Crapsi2007-L1544} observations.  Therefore, with 5\arcsec \, resolution, as for the integrated intensity map, the central flattening in the column density profile could not be detected, because it is only present within the central resolution elements of  \citet{Crapsi2007-L1544} observations.

We would also like to point out another important difference between our ALMA observations and those carried out by \citet{Crapsi2007-L1544}: our new ALMA image of p\nhtdoo \, is the result of a 3-point mosaic (see Fig.\,\ref{fig:L1544_mosaic}), while the IRAM-PdBI observations consist of a single-point image without including a primary beam correction. This may cause an artificially steeper flux drop away from the central region, mimicking a centrally concentrated pNH$_2$D flux profile consistent with the distribution of the dust continuum emission. Finally, this leads to the misleading conclusion that central freeze-out is not needed for deuterated ammonia, pointing out the importance of  high-angular resolution and high-sensitivity mosaic observations for a correct interpretation of pre-stellar core observations.


\section{NH$_3$(1,1) simulated spectra} \label{Appendix:nh3_spectra}
\begin{figure}
    \centering
    \includegraphics[width=0.7\textwidth,angle=0]{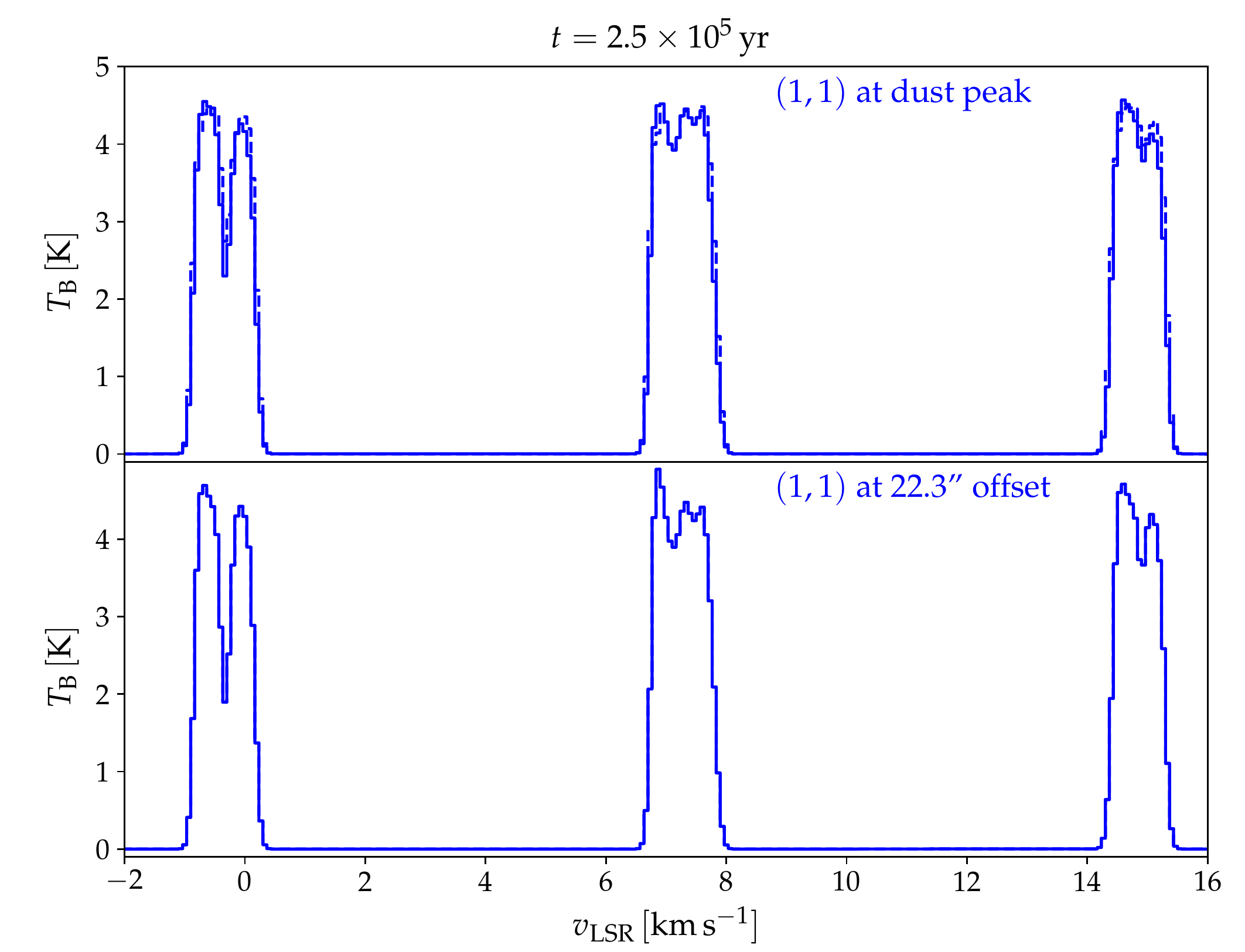}
    \caption{Simulated NH$_3$(1,1) spectra toward the dust peak (top panel) and an offset 22.3\arcsec \, (bottom panel) away from the center, to be compared with the two spectra in Figure 2 of \cite{Crapsi2007-L1544}. A beam of 4\arcsec \, \citep[close to the angular resolution of the VLA observations of][]{Crapsi2007-L1544} has been assumed. The solid blue histogram is the spectrum obtained with the NH$_3$ abundance profile predicted by our model and shown in Fig.\,\ref{fig:L1544_abu+fD}, while the dashed blue histogram (only visible in the top panel, as in the bottom panel it coincides with the solid histogram) is the spectrum predicted assuming NH$_3$ constant abundance within the central 5000\,au. Both spectra well reproduce the observed ones, within the observational uncertainties. It is then clear that NH$_3$(1,1) interferometric spectra of L1544 cannot provide constraints on the fractional abundance distribution of ammonia, unlike ALMA observations of \nhtdoo \, (see Fig.\,\ref{fig:spectra_nofreezeout}).}  
    \label{fig:nh3_spectra}
\end{figure}

\restartappendixnumbering
Given that previous work has claimed that the fractional abundance of NH$_3$ is actually increasing toward the L1544 center \citep[e.g.][]{Crapsi2007-L1544}, we would like to test if our model predictions about the NH$_3$ fractional abundance across the core (see Fig.\,\ref{fig:L1544_abu+fD}) is consistent with previous interferometric work.  For this, we simulate VLA NH$_3$(1,1) observations of  our model cloud using the LOC radiative transfer code and assuming a telescope beam of 4\arcsec \, \citep[close to the VLA synthesized beam in the observations of][ 4.34\arcsec $\times$ 3.45\arcsec]{Crapsi2007-L1544}, at two different positions: the dust peak and the offset (10\arcsec , -20\arcsec), as in Figure 2 of \citet{Crapsi2007-L1544}. The resultant spectra, focusing on the observed central three groups of hyperfine components, are displayed in Fig.\,\ref{fig:nh3_spectra} (solid histogram). Both spectra are in good agreement with the VLA observations, taking into account their 15\% flux uncertainty. Also, we are not simulating interferometric-only observations, as reported in \citet{Crapsi2007-L1544}, as we take into account all the flux from the various scales, including the extended envelope, filtered out by the VLA observations. This point will be made clear in Schmiedeke et al., in prep., where a higher-sensitivity map of NH$_3$(1,1) line obtained by combining VLA with GBT data will be presented. Therefore, we conclude that our model, where NH$_3$ presents a sharp drop in abundance toward the pre-stellar core center, is consistent with VLA observations. We also considered a case where the NH$_3$ fractional abundance does not drop in the central 6000\,au (unlike in our models, shown in Fig.\,\ref{fig:L1544_abu+fD}) and instead it maintains a constant value of 2$\times$10$^{-8}$. The resultant spectra are also in Fig.\,\ref{fig:nh3_spectra} (see dashed histograms); they are almost identical to the solid histograms, underlining the fact that NH$_3$(1,1) data cannot be used to constrain the abundance profile of NH$_3$ within the central region of the core. This is due to the relatively low critical density of NH$_3$(1,1) \citep[$\simeq4\times 10^{3}$\,cm$^{-3}$;][]{Maret2009}, which then makes the 1,1 inversion transition mainly sensitive to the outer parts of the pre-stellar core, hiding abundance variations in the central few thousand au. In a sense, this result is similar to that found for the p\nhtdoo \, line, with the important difference that p\nhtdoo \, has a significantly higher critical density than NH$_3$(1,1) and then it is more sensitive to variations in the \nhtd \, fractional abundance within the central regions (as demonstrated by Fig.\,\ref{fig:spectra_nofreezeout}). 

\end{document}